\documentclass[journal ]{new-aiaa}
\usepackage[utf8]{inputenc}
\usepackage{textcomp}

\usepackage{graphicx}
\usepackage{amsmath}
\usepackage[version=4]{mhchem}
\usepackage{siunitx}
\usepackage{longtable,tabularx}
\usepackage{overpic}

\newcommand{\press}{p}

\newcommand{\nv}{N}
\newcommand{\vjdex}{\mathsf{J}}

\newcommand{\state}{\mathsf{x}}
\newcommand{\joint}{\mathsf{z}}

\newcommand{\covarbase}{\Sigma}

\newcommand{\kalman}{K}
\newcommand{\kalmansvd}{\breve{K}}
\newcommand{\bias}{\mathsf{b}}

\newcommand{\latent}{\pmb{\xi}}
\newcommand{\data}{D}
\newcommand{\weights}{w}

\newcommand{\meanbase}{\mu}
\newcommand{\meanstate}{\meanbase_\state}
\newcommand{\meanmeas}{\meanbase_\meas}
\newcommand{\covar}[1]{\covarbase_{#1}}
\newcommand{\mean}[1]{\meanbase_{#1}}
\newcommand{\approxcovar}[1]{\hat{\covarbase}_{#1}}
\newcommand{\approxmean}[1]{\hat{\meanbase}_{#1}}

\newcommand{\statecovar}{\covarbase_{\state}}
\newcommand{\statestddev}{\sigma_{\state}}

\newcommand{\expect}{\mathbb{E}}

\newcommand{\Reals}{\mathbb{R}}
\newcommand{\statedim}{n}
\newcommand{\measdim}{d}
\newcommand{\rank}{r}
\newcommand{\staterank}{\rank_\state}
\newcommand{\measrank}{\rank_\meas}

\newcommand{\normaldist}{\mathcal{N}}
\newcommand{\probdist}{\pi}

\newcommand{\like}{L}

\newcommand{\observe}{h}
\newcommand{\invobserve}{g}
\newcommand{\observemat}{H}
\newcommand{\meas}{\mathsf{y}}

\newcommand{\meassvd}{\breve{\meas}}
\newcommand{\statesvd}{\breve{\state}}

\newcommand{\forecast}{f}
\newcommand{\timeindex}{k}
\newcommand{\forecastmat}{F}

\newcommand{\ensdim}{M}
\newcommand{\ensstate}{X}
\newcommand{\ensmeas}{Y}
\newcommand{\ensnoise}{\mathcal{E}}
\newcommand{\ensstatesvd}{\breve{\ensstate}}
\newcommand{\ensmeassvd}{\breve{\ensmeas}}
\newcommand{\ensnoisesvd}{\breve{\ensnoise}}

\newcommand{\ensdex}{i}
\newcommand{\statemember}[1]{\state^{#1}}

\newcommand{\noisemember}[1]{\noise^{#1}}
\newcommand{\pnoisemember}[1]{\pnoise^{#1}}

\newcommand{\ensones}{\mathsf{1}_{\ensdim}}

\newcommand{\truemeas}{\meas^{\ast}}
\newcommand{\truestate}{\state^{\ast}}
\newcommand{\noise}{\varepsilon}
\newcommand{\Noise}{\mathcal{E}}
\newcommand{\noisecovar}{\covarbase_{\Noise}}

\newcommand{\noisestddev}{\sigma_{\Noise}}

\newcommand{\pnoise}{\eta}
\newcommand{\pnoisecovar}{\covarbase_{\pnoise}}

\newcommand{\ident}{I}

\newcommand{\snapshot}{j}

\newcommand{\newstuff}[1]{\textcolor{black}{#1}}

\newcommand{\sect}{Section}

\title{A \newstuff{practical guide to} estimation and uncertainty quantification of aerodynamic flows}

\author{Jeff D. Eldredge \footnote{\begin{tabular}{p{2cm}p{12cm}}
   \raisebox{-1.5cm}{\includegraphics[width=2cm]{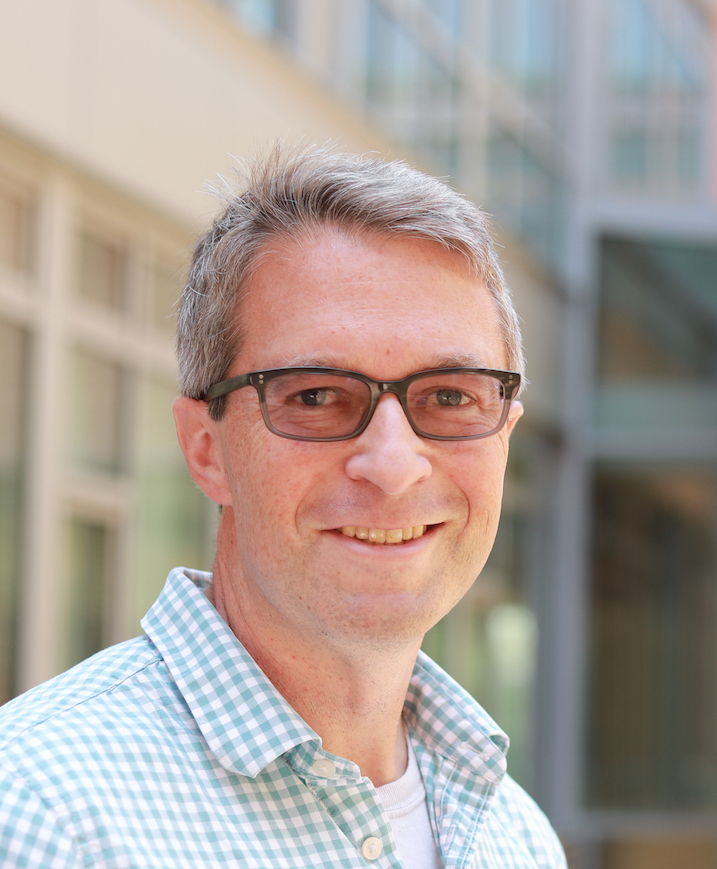}}   & Corresponding Author, \url{jdeldre@ucla.edu}.  Jeff Eldredge is Professor and Department Chair of Mechanical \& Aerospace Engineering at the University of California, Los Angeles, where has has served on the faculty since 2003. Prior to this, he was a post-doctoral researcher in Engineering at Cambridge University, and completed his Ph.D. at Caltech and his B.S. at Cornell University. He is a Fellow of the American Physical Society and an Associate Fellow of AIAA.
\end{tabular}} and Hanieh Mousavi\footnote{\begin{tabular}{p{2cm}p{12cm}}
   \raisebox{-1.5cm}{\includegraphics[width=2cm]{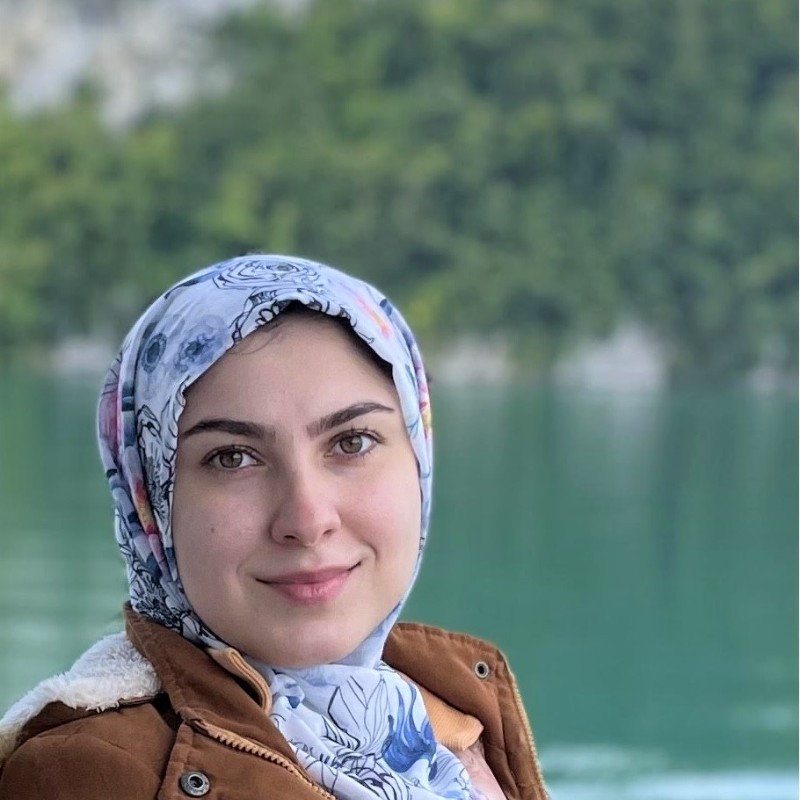}}   & Hanieh Mousavi is a graduate student in the Department of Mechanical \& Aerospace Engineering at the University of California, Los Angeles. Her research focuses on data assimilation in unsteady and disturbed aerodynamics.
\end{tabular}}}

\affil{Mechanical \& Aerospace Engineering, University of California, Los Angeles, Los Angeles, CA, 90095}

\begin{document}

\maketitle

\begin{abstract}
Many applications in aerodynamics, particularly in closed-loop control, depend on sensors to estimate the evolving state of the flow. This estimation task is inherently accompanied by uncertainty due to the noisy measurements of sensors or the non-uniqueness of the underlying mapping. Knowledge of this uncertainty can be as important for decision-making as that of the state itself. Uncertainty tracking is challenged by the often-nonlinear relationship between the measurements and the flow state. For example, a collection of passing vortices leaves a footprint in wall pressure that depends nonlinearly on the vortices' strengths and positions. In this paper, we outline recent approaches to flow estimation and illuminate them with worked examples and selected case studies. We review relevant probability tools, including sampling and estimation, in the powerful setting of Bayesian inference and demonstrate these in static flow estimation examples. We then review unsteady examples and illustrate the application of sequential estimation, and particularly, the ensemble Kalman filter. Finally, we discuss uncertainty quantification in neural network approximations of the mappings between sensor measurements and flow states. Recent aerodynamic applications have shown that the flow state can be encoded into a very low-dimensional latent space. We discuss the uncertainty implications of this encoding.
\end{abstract}

\section{Introduction}\label{Introduction}

\lettrine{T}{he} focus of this paper is on addressing problems in which we seek to estimate the state of a fluid flow from a set of sensor measurements. Though we will not address the purpose of this flow estimation, it ostensibly fits in with some larger task of flow or flight control: We would like to act to achieve some objective (maintain steady forces, perform agile maneuvers) in the presence of atmospheric disturbances (i.e., gusts) \cite{jones2022physics}. The signature of these disturbances can be found in the sensor data, and we would like to use these data to form a more complete picture of the flow.  

What constitutes a complete picture depends on the circumstances. In this paper, the {\em state}, $\state$, we seek will generally be the full flow field (velocity or vorticity). The actual flow field is infinite-dimensional, so in practice we seek a finite-dimensional representation of this field, e.g., the velocity on a grid of points, the coefficients of a modal expansion, or the positions and strengths of a set of vortex elements. However, we might not be so ambitious, and the state we seek might simply comprise the forces and moments on a wing or blade. It can also include other information we need but are not sure about, such as the parameterized geometry of a bounding surface, the Reynolds number, or the leading-edge suction parameter. Collectively, we can take this state to be a vector in a real-valued $\statedim$-dimensional space, $\state \in \Reals^\statedim$. We seek to infer this state from the finite-dimensional observation vector, $\meas \in \Reals^\measdim$, representing a set of $\measdim$ sensor measurements taken at some instant. The measurements $\meas$ might comprise data from pressure sensors, a force and moment sensor, an inertial measurement unit attached to a body in the flow, a hotwire measurement, particle image velocimetry data, Doppler ultrasound, or LIDAR measurements.

We will assume that there is a (possibly nonlinear) function $\observe$ that uniquely maps $\state$ to $\meas$,
\begin{equation}
\label{eq:observe}
    \meas = \observe(\state),
\end{equation}
called the {\em observation} operator. In other words, given $\state$ and the function $\observe$, we have enough information to calculate $\meas$, our prediction of the measurements associated with this state. Before we start, we should recognize that, just as the state $\state$ is a finite-dimensional representation of the actual infinite-dimensional state, the operator $\observe$ is only an approximation of the actual observation operator.

There are a number of data-oriented goals regarding $\observe$ that may interest us in aerodynamics. A few key goals are the following:
\begin{enumerate}
    \item {\em Inversion}: We know $\observe$ well (e.g., via physics) and can evaluate it at any $\state$, but we are interested in finding $\state$ given $\meas$. Roughly speaking, we seek $\state = \observe^{-1}(\meas)$. Note that,  although the observation operator maps $\state$ uniquely to $\meas$, we don't know whether this inverse maps $\meas$ to a unique $\state$. We also expect there to be an error because $\observe$ is simplified in some fashion.  For example, for pressure measurements, $\observe$ might involve a grid-based solution of the pressure Poisson equation or an evaluation of the Bernoulli equation. If sensors measure velocity, then $\observe$ might obtain this directly from the state vector (e.g., by interpolation). 
    \item {\em Regression}: We don’t know $\observe$ (or only partially know it), but we have a set of training data, comprising pairs $(\state_\snapshot,\meas_\snapshot)$, and we wish to learn an approximation for the true observation operator $\tilde{\observe}(\state; \weights) \approx \observe(\state)$ or its inverse $\tilde{\invobserve}(\meas; \weights) \approx \observe^{-1}(\meas)$, e.g., using a neural network parameterized by a set of weights and biases, $\weights$. 
\end{enumerate}
We might also seek to accomplish both goals, i.e., obtain weights $\weights$ and then use the approximate operator $\tilde{\observe}(\state; w)$ for inversion.

Regardless of whether $\observe$ is known or derived from data, our focus in this paper will be on the inversion goal. It is important to note that the measurements $\meas$ from sensors are inherently noisy, and we will have to contend with this noise in the inversion task. In fact, this noise is an opportunity to approach the problem {\em probabilistically}, and the tools of probability---and particularly, Bayesian probability---empower us to fully characterize the uncertainty of the inferred state and address a number of probing questions: About which components are we most uncertain? Which sensors are most informative? Are there multiple probable states associated with these measurements? Thus, even if the measurements are not particularly noisy, it is nonetheless useful to frame the inversion problem in this manner. We will address this Bayesian inversion problem in \sect~\ref{sec:static_inference}, and use the opportunity to review basic concepts and tools.  

In an unsteady flow, it is crucial to remember that the state is evolving in time, so there is an underlying dynamical (or {\em forecast}) operator $\forecast_\timeindex$ that updates the state from one time to the next,
\begin{equation}
\label{eq:forecast}
        \state_{\timeindex} = \forecast_\timeindex\left(\state_{\timeindex-1} \right),
\end{equation}
where $\timeindex$ denotes the index of the current time, $t_\timeindex$. 
As with $\observe$, we may know $\forecast$ from physics or from regression over training data. Since $\state$ represents the flow state, then $\forecast$ is likely to be some finite-dimensional representation of the Navier-Stokes or Euler equations. If $\state$ contains grid velocity or vorticity data, then $\forecast$ might comprise one time step of a CFD solution. If $\state$ are modal expansion coefficients, then $\forecast$ could be a Galerkin projection of Navier-Stokes onto the modes. If $\state$ comprises vortex element parameters, then $\forecast$ might be the vorticity transport equation in Lagrangian form. And $\forecast$ might be a neural operator $\tilde{\forecast}(\state;\weights)$ learned from data, in the same way that $\observe$ might. In these unsteady problems, we stress that \eqref{eq:observe} represents an instantaneous relationship between state and measurements. Our goal in an unsteady problem is to fold these measurements into our evolving prediction of the state, ostensibly with the goal of improving this prediction.    

How do we fold these together? This is the central objective of data assimilation, which offers a few different approaches to the problem. However, if we acknowledge that the forecast, like the measurements, has inherent uncertainty, then the probabilistic approach enables us to blend the forecast and the measurement update by balancing the trust between them, resulting in a sequential estimator of the state. As we will show in \sect~\ref{sec:sequential_inference}, each measurement update is a Bayesian inversion that starts from the prior obtained from the most recent forecast step, so sequential estimation draws heavily from the tools in \sect~\ref{sec:static_inference}. We will finish the paper with a discussion of uncertainty quantification using learned operators, i.e., those obtained by regression over training data, in \sect~\ref{sec:neural_network}. But first, we will review necessary concepts and tools from Bayesian probability theory.

\newstuff{We apologize preemptively that this paper will fall well short of a comprehensive review of the recent work on the vibrant field of flow estimation. Our objective here is to provide an overview and brief tutorial on state-of-the-art tools and to provide selected examples from the literature that illustrate these tools.}

\section{Static inference of flows}\label{sec:static_inference}

In this section we will address the problem of inverting \eqref{eq:observe} in a probabilistic sense. That is, we will think of the state $\state$ and observations $\meas$ as continuous random variables, and ask the question: given $\meas$, what is the most likely $\state$? To address this question, we will review several concepts and tools from probability theory. This review is meant to be cursory and illustrative, and assumes some basic knowledge of probability. For more detailed coverage, the reader is referred to any of several books on the subject. We particularly recommend the books by Bishop \cite{bishop2006pattern} and Asch et al.~\cite{asch2016data}, both of which have one or more chapters devoted to basics of Bayesian probability. We also highly recommend the reviews by Stuart~\cite{stuart2010inverse} \newstuff{and by Ghattas and Willcox \cite{ghattas2021learning}, the latter of which is devoted to addressing the dovetailing tasks of inverse problems and model reduction. We also direct the reader's attention to the excellent book by Tarantola \cite{tarantola2005inverse}, which provides more depth to the probabilistic setting of the inverse problem.}

The key concept that encapsulates the answer to the question above (and several other related ones) is the conditional probability distribution,
\begin{equation}
\label{eq:condprob}
    \probdist(\state | \meas),
\end{equation}
which describes the distribution of possible states $\state$, given a particular observation $\meas$. If we knew \eqref{eq:condprob}, then we would have all the information we need about the inversion problem, and we would access that information through conditional expectations: that is, if we wish to know the expected value of some function of the state, $F(\state)$ based on the available measurement, we would compute
\begin{equation}
\label{eq:expect}
    \expect(F|\meas) = \int F(\state) \probdist(\state|\meas)\,\mathrm{d}\state, 
\end{equation}
which sensibly puts more weight on more likely states. Note that the function $F(\state)$ could be just $\state$ itself, in which case \eqref{eq:expect} returns the expected state (the conditional mean). One could also, for example, compute the expected state covariance to assess our uncertainty of the expected state. Equations \eqref{eq:meanx}--\eqref{eq:sigyy} in Appendix \ref{sec:appendix-linear} summarize the most important quantities.

\subsection{The Bayesian view of inversion}

How do we get the conditional probability distribution \eqref{eq:condprob}? First, it is useful to remember that we have access to the observation operator \eqref{eq:observe}, and we will show later that this means that, after we account for measurement noise, we know the opposite conditional distribution, the distribution of possible observations at a given state. This is called the {\em likelihood function}, and we will denote it by $\like(\meas|\state)$. To get the conditional we seek, we turn to some simple but very important identities for the joint probability distribution, $\probdist(\state,\meas)$. This distribution measures the probability of getting a particular pair of values of the state and the observation. We can assemble this joint distribution in two different ways, each by multiplying a marginal distribution by a conditional:
\begin{equation}
\label{eq:bayes0}
    \probdist(\state,\meas) = \probdist(\state|\meas)\probdist(\meas) = \like(\meas|\state) \probdist_0(\state).
\end{equation}
We arrive at Bayes' theorem by solving for one of these conditionals (the one we want), in terms of the others:
\begin{equation}
\label{eq:bayes1}
   \probdist(\state|\meas) = \frac{\like(\meas|\state) \probdist_0(\state)}{\probdist(\meas)}.
\end{equation}
This theorem is simple but provides a very powerful way of framing the inversion problem in a probabilistic manner \cite{stuart2010inverse}: We start with some basic knowledge of the distribution of possible states $\probdist_0(\state)$, called the {\em prior} distribution. Our objective is to tighten our knowledge of $\state$ by obtaining the {\em posterior} distribution, $\probdist(\state|\meas)$, i.e., the distribution of states conditioned on a measurement. The connection between the posterior and prior is through the likelihood. The denominator, $\probdist(\meas)$, is a marginal distribution of the possible measurement values integrated across all possible states. But in using Bayes' theorem to obtain the posterior distribution, it is evaluated at a particular value of the observation vector---the true measurement vector, $\truemeas$---and remains only a function of $\state$, so $\probdist(\truemeas)$ is simply a uniform scaling factor. Often its value is not important, since we only care about comparing the probabilities of two states, and we are content to know the following:
\begin{equation}
\label{eq:bayes2}
    \probdist(\state|\truemeas) \ \propto \ \like(\truemeas|\state) \ \probdist_0(\state).
\end{equation}

The likelihood function is obviously key, and it follows directly from the observation operator and knowledge of the type of measurement noise we expect. Accounting for this noise, $\noise$, the observation model is
\begin{equation}
\label{eq:observenoise}
    \meas = \observe(\state) + \noise.
\end{equation}
Often it is safe to assume that this noise is Gaussian with zero mean: that is, $\noise \sim \normaldist(0,\noisecovar)$, where the covariance $\noisecovar \in \Reals^{\measdim\times\measdim}$ is diagonal (i.e., the sensors' noise is independent of one another). Equation~\eqref{eq:observenoise} then implies that the measurements are also Gaussian distributed, about $\observe(\state)$, i.e., $\like(\meas|\state) =  \normaldist(\meas|\observe(\state),\noisecovar)$, indicating that we trust our predicted observation only to within the measurement noise. Multiplying this likelihood with the prior gives us access to the posterior distribution.\footnote{\newstuff{The likelihood's evaluation at the true measurement could be thought of as a product of two distributions: an exact distribution representing the (perfect) observation model, and a distribution about the true measurement representing sensor noise. This is the view described by Tarantola \cite{tarantola2005inverse}.}}

\begin{figure}
    \centering
    \begin{overpic}[width=0.46\linewidth]{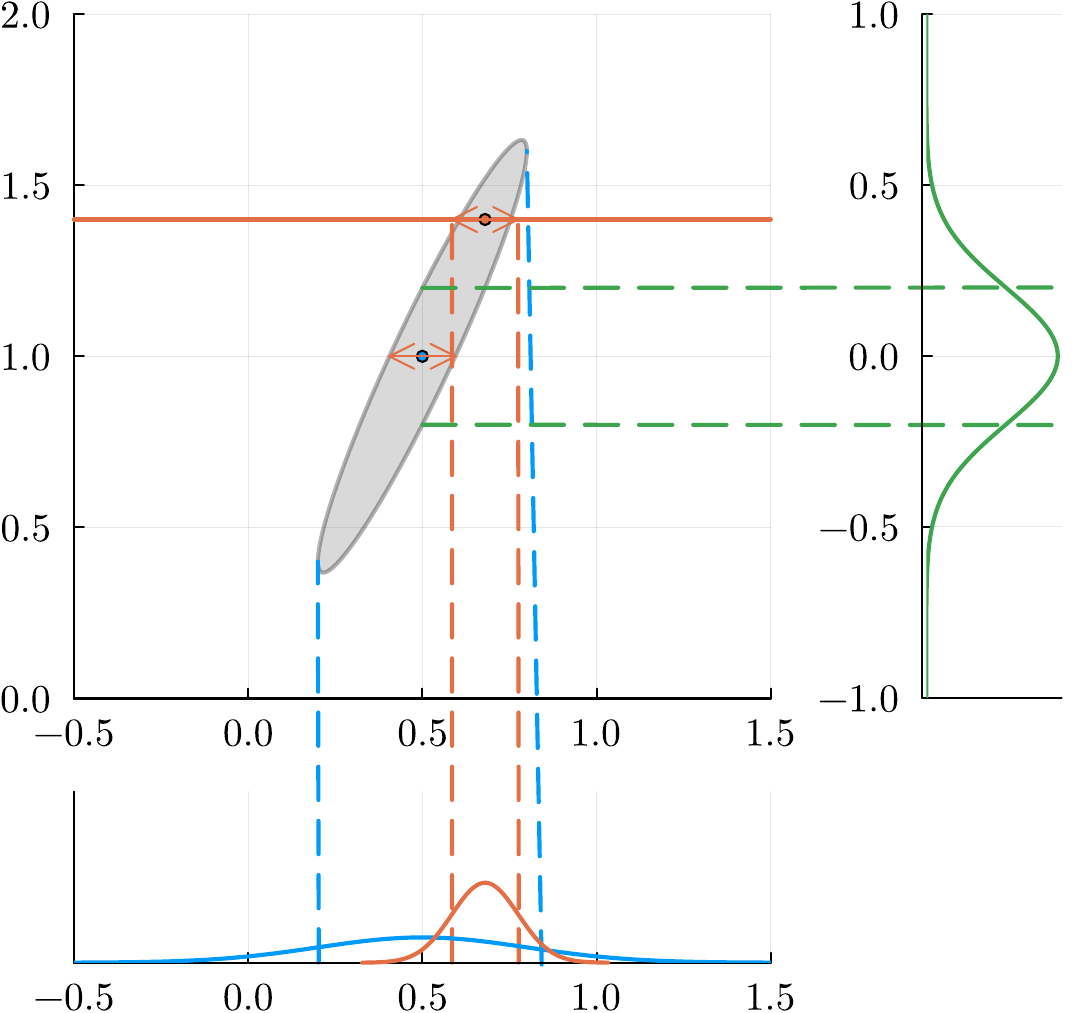}
    \put(-10,90){(a)}
    \put(40,20){\footnotesize $\state$}
    \put(-5,60){\rotatebox{90}{\footnotesize $\meas$}}
    \put(40,-4){\footnotesize $\state$}
    \put(75,60){\rotatebox{90}{\footnotesize $\noise$}}
    \put(20,7){\footnotesize $\probdist_0(\state)$}
    \put(50,13){\footnotesize $\probdist(\state|\truemeas)$}
    \put(0,72){\footnotesize $\truemeas$}
    \end{overpic}\qquad
    \begin{overpic}[width=0.46\linewidth]{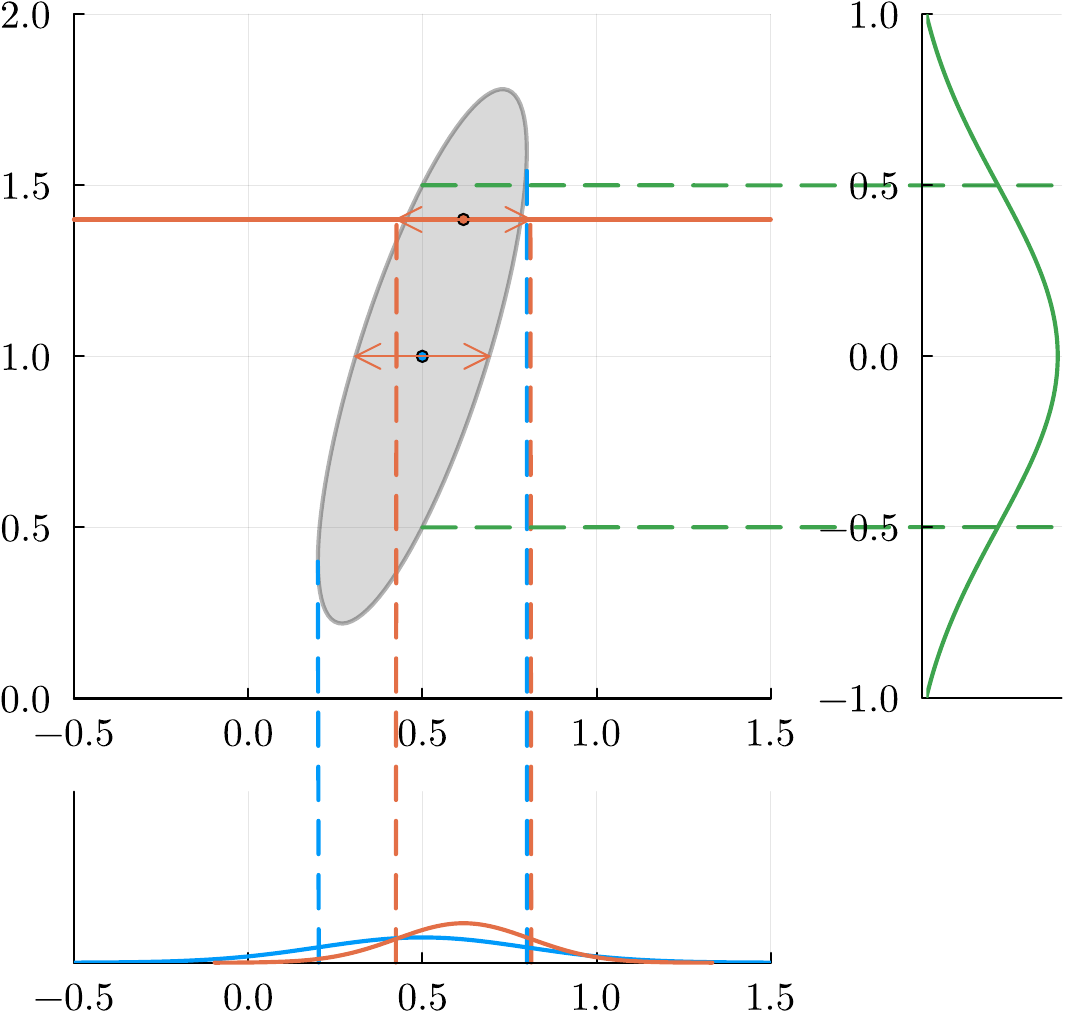}
    \put(-10,90){(b)}
    \put(40,20){\footnotesize $\state$}
    \put(-5,60){\rotatebox{90}{\footnotesize $\meas$}}
    \put(40,-4){\footnotesize $\state$}
    \put(75,60){\rotatebox{90}{\footnotesize $\noise$}}
    \put(20,7){\footnotesize $\probdist_0(\state)$}
    \put(50,8){\footnotesize $\probdist(\state|\truemeas)$}
    \put(0,72){\footnotesize $\truemeas$}
    \end{overpic}
    \caption{Construction of the posterior distribution (red) for observation model $\observe(\state) = 2\state$ and Gaussian prior (blue) with varying degrees of measurement noise (green) and a true measurement $\truemeas = 1.4$. Covariance ellipse of the joint distribution shown in gray. (a) $\noisestddev = 0.1$, (b) $\noisestddev = 0.5$}
    \label{fig:bayes-example}
\end{figure}

Bayes' theorem has been used by numerous investigators to frame ill-posed inverse problems in a probabilistic setting. As an example in the context of fluid dynamics, Kontogiannis et al.~\cite{kontogiannis2022joint} have recently used a Bayesian framework to assimilate noisy images of a flow field (from magnetic resonance velocimetry) to reconstruct the flow through a channel and infer the channel's geometry. The prior enables them to restrict the estimated flow to solutions of the Navier--Stokes equations. The likelihood expresses the error between the noisy velocity measurements and the velocity computed on a mesh and interpolated to the same locations as the measurements.

We can illustrate the idea of the Bayesian approach to inversion with a simple example. Suppose that $\state$ and $\meas$ are both scalar-valued, and the observation operator is linear, $\observe(\state) = 2\state$. We will assume the prior is Gaussian with mean $\meanstate = 0.5$ and standard deviation $\statestddev = 0.3$. Figure~\ref{fig:bayes-example} depicts the resulting posterior distribution for two different levels of measurement noise when the true measurement is $\truemeas = 1.4$. The joint distribution, shown as a gray covariance ellipse, is assembled from the product of the prior (which sets the ellipse's horizontal extent) and the likelihood (the vertical width, determined by the noise distribution, shown in green). The posterior mean then lies at the center of the intersection between the true measurement (a red horizontal line) and the ellipse, and the posterior covariance---the uncertainty of our new estimate---is set by the maximum horizontal width of the covariance ellipse (the red arrow).  In one case, the measurement noise is low ($\noisestddev = 0.1$), and the resulting posterior is much narrower than the prior and centered at a mean of nearly $0.7$, the value we would get if we inverted without noise. But with larger measurement noise ($\noisestddev = 0.4$), the true measurement is trusted less, so the posterior relies more on the prior, lying just to the right of it and only slightly narrower. 

\subsection{Linear observation operators and Gaussian distributions}

This simple scalar example with a linear observation operator and Gaussian noise and prior distributions extends nicely to multidimensional problems, in which the observation operator is a matrix, $\observe(\state) = \observemat\state$. This extension follows because products of multivariate Gaussians are also Gaussian \cite{bishop2006pattern}, so, if the prior is $\probdist_0(\state) = \normaldist(\state|\meanstate, \statecovar)$, then the posterior is also Gaussian, $\probdist(\state|\truemeas) = \normaldist(\state|\mean{\state|\truemeas}, \covar{\state|\truemeas})$, where the posterior mean and covariance are (see \ref{sec:appendix-linear})
\begin{align}
    \mean{\state|\truemeas} &= \meanstate + \kalman \left( \truemeas - \observemat \meanstate\right),\label{eq:kalmanupdate-mean-linear}\\ \covar{\state|\truemeas} & = \left(\statecovar^{-1} + \observemat^T \covar{\Noise}^{-1} \observemat  \right)^{-1} = \left(\ident_\statedim - \kalman \observemat \right) \statecovar, \label{eq:kalmanupdate-covar-linear}
\end{align} 
and where the matrix $\kalman$ is given by
\begin{equation}
\label{eq:kalman2}
    \kalman = \statecovar \observemat^T \left( \covar{\Noise} + \observemat \statecovar^{-1} \observemat^T\right)^{-1}.
\end{equation}
The second of the two identities in \eqref{eq:kalmanupdate-covar-linear} follows from the Woodbury matrix identity. 

The mean and covariance in \eqref{eq:kalmanupdate-mean-linear} and \eqref{eq:kalmanupdate-covar-linear} give us complete information about the state---its expected value and its uncertainty---based on the measurements. The matrix $\kalman$, the Kalman gain, has a particularly important role in data assimilation. It acts as a gain on the innovation (the difference $\truemeas - \observemat \meanstate$ between the true measurement and the measurement we predict based on the prior) to shift the prior mean to the posterior; it also shrinks the prior covariance to its posterior value. The Woodbury matrix identity can be used to rewrite $\kalman$ in another illuminating form,
\begin{equation}
\label{eq:kalman}
    \kalman = \left(\statecovar^{-1} + \observemat^T \covar{\Noise}^{-1} \observemat \right)^{-1} \observemat^T \covar{\Noise}^{-1}.
\end{equation}
When the measurement noise is uniform and goes to zero (or the prior uncertainty is large), the Kalman gain in \eqref{eq:kalman} reduces to the pseudo-inverse, $(\observemat^T\observemat)^{-1}\observemat^T$, and the posterior covariance in \eqref{eq:kalmanupdate-covar-linear} to $(\observemat^T \covar{\Noise}^{-1} \observemat)^{-1}$, indicating full trust in the measurements over the prior. In contrast, when the measurements are very noisy (and untrustworthy), then the Kalman gain itself shrinks to zero, keeping the posterior near the prior.

\paragraph{A simple fluid dynamics example.} In multi-dimensional problems, the Kalman gain can be unpacked to reveal a lot of interesting detail about how the measurements inform the inference of the state. We will use a very simple fluid dynamics task to demonstrate this. Suppose we have a U-tube manometer for measuring pressure in a test section relative to the ambient, $\Delta p = p - p_\infty$. However, we wish to use it to infer the absolute pressures, $p$ and $p_\infty$. Clearly this is impossible, but it is illustrative to see how this impossible task fails. Our measurement vector is one dimensional, $\meas = \Delta p$, our state vector is two dimensional, $\state = [p\,\,\,p_\infty]^T$, and the observation operator is the $1\times 2$ matrix $\observemat = [1\,\,\, -1]$. We will assume that the ambient pressure has historically been within 100 Pa of 1 atm ($101325$ Pa), and we do not have any further information about the pressure in the test section, so we will set the prior to be Gaussian, with mean and covariance
\begin{equation}
    \meanstate = 101325 \begin{pmatrix}1 \\ 1\end{pmatrix}, \qquad \statecovar = 100^2\begin{bmatrix} 1 & 0 \\ 0 & 1 \end{bmatrix}.
\end{equation}   
We will assume that our U-tube can measure pressure with a standard deviation of $\noisestddev = 10$ Pa.

\begin{figure}
    \centering
    \begin{overpic}[width=0.46\linewidth]{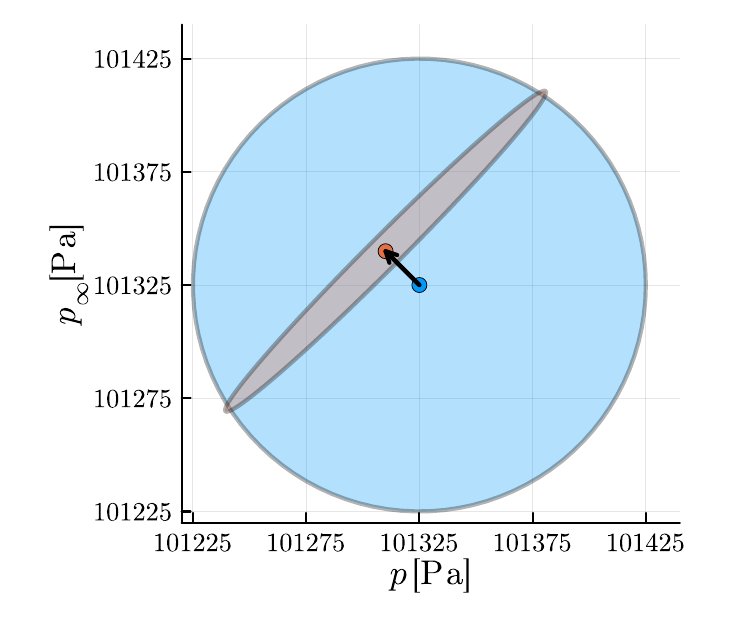}
    \put(55,50){$v_1$}
    \end{overpic}
    \caption{Prior (blue) and posterior (gray) covariance ellipses in the state space of the manometer example, with the respective means shown as small circles. The informative vector $v_1$ is also shown.}
    \label{fig:manometer-covar}
\end{figure}

Figure~\ref{fig:manometer-covar} depicts the posterior mean and covariance that results when we take a manometer measurement of $\truemeas = -30$ Pa. Whereas the prior distribution had a mean of zero and circular covariance because of our identical uncertainty about both $p$ and $p_\infty$, the new mean is at $\mean{\state|\truemeas} = [101310\,\,\,101340]^T$, whose components differ by $-30$, and the new covariance ellipse is still as broad as the prior circle in one direction but much narrower in the perpendicular direction. How can we understand this result? We will address this question by decomposing the observation matrix $\observemat$ with a singular value decomposition (SVD); to normalize the state and observation vectors, we will also re-scale the matrix with the noise and prior covariances,
\begin{equation}
\label{eq:Hdecompose}
    \observemat = \covar{\Noise}^{1/2} U S V^T \statecovar^{-1/2}.
\end{equation}
The columns in $U$ and $V$ form orthonormal bases for the measurement and the state spaces, respectively, and $S$ is a diagonal matrix of singular values. (We use the full SVD here, so $S\in \Reals^{\measdim\times\statedim}$, not necessarily square.) We can write $\meas$ and $\state$ in these bases (and ``whiten'' them with the covariances so they have comparable components in these bases):
\begin{equation}
    \meas = \covar{\Noise}^{1/2} U \meassvd, \quad \state = \statecovar^{1/2} V \statesvd.
\end{equation}
By virtue of these new bases, the observation operator reduces to the completely decoupled form,
\begin{equation}
\label{eq:observe-svd}
    \meassvd = S \statesvd.
\end{equation}
The covariances can all be decomposed accordingly, and are shown in \eqref{eq:covar-svd}.

In the current example,
\begin{equation}
    U = \begin{bmatrix}
        -1
    \end{bmatrix}, \quad
    S = \begin{bmatrix}
        14.14 & 0
    \end{bmatrix},\quad
    V = \begin{bmatrix}
        -1/\sqrt{2} & 1/\sqrt{2} \\
        1/\sqrt{2} & 1/\sqrt{2}
    \end{bmatrix}.
\end{equation}
The first column of $V$, the vector $v_1$, represents differences in the entries in $\state$, reflecting the manometer's function. In \eqref{eq:observe-svd}, the corresponding component of $\statesvd$ is multiplied by the only non-zero singular value to produce the sole component of $\meassvd$, and from that, the predicted measurement $\meas$. This first column of $V$ thus spans the informative subspace of $\observemat$. In contrast, the second column of $V$ (vector $v_2$) represents the sum of the entries in $\state$ (i.e., the average of the pressures), and this component is multiplied by zero. In other words, the average of the pressures lies in the null space of $\observemat$, since the manometer has no ability to sense this average; this second column spans the noninformative, or non-observable, subspace of the state space. In general, when the number of sensors is less than the number of states, the observation matrix is rank deficient: $S$ amplifies some of the state components while diminishing or completely removing others. Those that are removed are not observable by the sensors; the remaining ones are observable to varying degrees. \newstuff{Examples 2.1 and 2.2 by Stuart \cite{stuart2010inverse} provide nice illustrations of under- and overdetermined systems.}

This decomposition \eqref{eq:Hdecompose} of the observation operator shows us the degree to which the sensors are impacted by the state. In turn, it can also show us how the sensors inform the state correction \eqref{eq:kalmanupdate-mean-linear} via $\covar{\state|\truemeas}$ and $\kalman$, because these two matrices are also nicely factorized into modes by the SVD of $\observemat$ and follow the same relations with the covariances in the new bases as they do in the original bases. The posterior covariance is factorized as
\begin{equation}
\label{eq:post-covar-svd}
    \covar{\state|\truemeas} = \statecovar^{1/2} V  \covar{\statesvd|\truemeas} V^T \statecovar^{1/2},
\end{equation}
where
\begin{equation}
\label{eq:covar-svd-def}
    \covar{\statesvd|\truemeas} \equiv \covar{\statesvd} - \covar{\statesvd\meassvd}\covar{\meassvd\meassvd}^{-1}\covar{\statesvd\meassvd}^T \equiv \left(\ident_\statedim + S^T S\right)^{-1};
\end{equation}
and the Kalman gain as
\begin{equation}
\label{eq:kalman-svd}
    \kalman = \statecovar^{1/2} V \kalmansvd  U^T \noisecovar^{-1/2},
\end{equation}
where
\begin{equation}
\label{eq:kalman-svd-def}
    \kalmansvd \equiv \covar{\statesvd\meassvd} \covar{\meassvd\meassvd}^{-1} \equiv S^T \left(\ident_\measdim + S S^T\right)^{-1} .
\end{equation}
The diagonal $\statedim \times \statedim$ posterior covariance matrix $\covar{\statesvd|\truemeas} = \left(\ident_\statedim  + S^T S\right)^{-1}$ has entries that are less than 1 where the singular values in $S$ are nonzero, and equal to 1 where the singular values vanish. This has the effect of reducing the covariance in \eqref{eq:post-covar-svd} only in the observable directions of $V$, and leaving the non-observable directions the same as in the prior. This is easily seen in Figure~\ref{fig:manometer-covar}, in which the uncertainty in the observable direction $v_1 = (-1/\sqrt{2},1/\sqrt{2})$ (associated with pressure differences) is shrunk, but the non-observable direction $v_2 = (1/\sqrt{2},1/\sqrt{2})$ (associated with average pressures) is untouched.

The decomposed form of the Kalman gain \eqref{eq:kalman-svd} reveals that the first operation that $\kalman$ applies to the measurement innovation in \eqref{eq:kalmanupdate-mean-linear} is to scale it by $\noisecovar^{1/2}$: sensors with larger noise are thereby assigned lower values in this whitened form. The scaled innovation is then rotated into a new coordinate system by $U^T$, in which the modal Kalman gain $\kalmansvd$ can act upon each component individually. In equation \eqref{eq:kalman-svd-def}, the zero entries in the matrix $S^T$ ensure that the non-observable components of the state are not updated by the sensors. In the manometer example, the mean is shifted only in the informative direction $(-1/\sqrt{2},1/\sqrt{2})$.

This example has illuminated the role of $V$ as a useful basis for the state space, distinguishing observable from non-observable directions in the state space. What meaning can we attach to $U$? With only one sensor, it is difficult to illustrate its role, but suppose we add a barometer to measure the ambient pressure, so that $\measdim = 2$. We now have a pair of two-dimensional column vectors in $U$, two singular values, and $\observemat$ is no longer rank degenerate. However, if this barometer's measurements are very noisy compared to the manometer, then the second singular value would be much smaller than the first, and the first column in $U$ would place much more weight on the manometer than the barometer. Thus, the singular values express the signal-to-noise ratio of the sensors, and the column vectors of $U$ weight the measurements accordingly so that the state update is not corrupted by measurement noise.

We note in passing that we could also have compensated for the missing direction in the state update by endowing the prior with additional knowledge of the ambient pressure, say, through a previous (less noisy) barometer measurement. We would set both entries in the prior mean to this ambient pressure measurement, and set the ambient pressure's diagonal entry in the prior covariance to a much smaller value to reflect the barometer's uncertainty. The resulting posterior mean would then preserve the value of the second component (ambient pressure), and estimate the test section pressure to be approximately 30 Pa smaller. In the absence of information from the sensors, the prior is the fallback. This approach also demonstrates that Bayesian inversion can be used successively with a sequence of measurements, with each posterior from one measurement serving as the prior for the next. Indeed, this successive use will be important when we discuss sequential estimation in Section~\ref{sec:sequential_inference}.

\subsection{Nonlinear observation operators}

Unfortunately in aerodynamics, it is rare that the observation operator is linear, and we frequently encounter cases in which the measurements depend nonlinearly on the state. An obvious example is pressure, which depends quadratically on the velocity (or vorticity) in inertially-dominated flows. The primary challenge with nonlinear observation operators is that, even if the prior and measurement noise are Gaussian, the likelihood (and thus, the posterior) is not Gaussian in the state. Revisiting \eqref{eq:bayes2} and \eqref{eq:observenoise}, we recall that the likelihood involves additive Gaussian noise about a mean given by $\observe(\state)$. The product of the likelihood and prior gives rise to an exponential whose argument is the negative of
\begin{equation}
\label{eq:posterior-arg}
    \frac{1}{2}(\truemeas - \observe(\state))^T \noisecovar^{-1} (\truemeas - \observe(\state)) + \frac{1}{2} (\state-\meanstate)^T \statecovar^{-1} (\state-\meanstate),
\end{equation}
which cannot generally be reduced to a quadratic form for $\state$, as we would need for a Gaussian posterior. Thus, we cannot readily apply the tools of the previous section to find the posterior distribution analytically (though, as we will show, these tools remain valuable). We also cannot analytically compute the marginal distribution in the denominator of \eqref{eq:bayes1}, if we need it.

\subsubsection{Sampling the posterior}
So how do we proceed? If our goal is just to find the expected state given the true measurement, then we should note that this is the $\state$ that minimizes \eqref{eq:posterior-arg}. This note effectively frames the state estimation as an optimization problem, called the {\em maximum a posteriori (MAP)} estimate, and we could use optimization tools to solve it. However, our goal is also to characterize the posterior uncertainty, so we still seek the posterior distribution in some form. We have a few options. If our prior distribution is already reasonably precise, then we do not have to move very far from it to obtain the posterior. In such a case we can linearize the observation operator about the prior mean and then use the Kalman update that we presented in the previous section, taking $\observemat \approx \nabla\observe(\meanstate)$. We will revisit this idea later when we discuss sequential estimation. However, if our prior has a lot of uncertainty, or we suspect that the solution $\state$ to \eqref{eq:observe} is not unique, then we need a more thorough strategy, and we can get this by {\em sampling} the posterior. Sampling from a distribution provides us with an indirect tool for characterizing it: with an ensemble $\{\statemember{\ensdex}\}_{\ensdex=1}^{\ensdim}$ of $\ensdim$ independent samples, each drawn from the posterior distribution $\statemember{\ensdex} \sim \probdist(\state|\truemeas)$, we can approximately compute expected values by summing over the samples,
\begin{equation}
    \expect[F|\truemeas] = \int F(\state) \probdist(\state|\truemeas) \,\mathrm{d}\state \approx \frac{1}{\ensdim}\sum_{\ensdex=1}^\ensdim F(\statemember{\ensdex}).
\end{equation}
We can also use the samples to develop approximate probabilistic models of the posterior, $\tilde{\pi}(\state|\meas)$, to do further work, e.g., to serve as a generative model for drawing new instances of the state. There are many approaches for carrying out this approximation task, most of which consist of minimizing (over the available data) the ``distance'' between the approximate probability distribution and the true one; in probability theory, the distance between two distributions is the so-called Kullback-Leibler (KL) divergence \cite{kullback1951information}. Various families of approximate distributions can be used, e.g., mixture models \cite{bishop2006pattern}. 

We have ready access to tools to draw samples from simple distributions: e.g., the \verb+randn+ function in many programming languages generates samples of a one-dimensional Gaussian distribution with desired parameters, and we can create a histogram of the samples to verify this. However, for more exotic distributions in higher-dimensional spaces, we need an algorithm. A naive approach would be a grid search: we systematically evaluate the probability at every sample point on an $\statedim$-dimensional grid of points. However, this would waste most of its effort on areas with low probability, and this wastefulness gets worse as $\statedim$ increases---the so-called curse of dimensionality. Instead, we can rely on Markov chain Monte Carlo (MCMC), a class of methods that construct a chain of samples that concentrate in the high-probability regions and approximate a desired distribution. (We also typically only keep every $L$th chain entry to ensure the samples are independent.) A feature of MCMC that makes it particularly helpful is that it only compares (i.e., considers the ratio of) the probability of a candidate sample point with that of a previous one to decide whether to accept it in the chain. Since the probability distribution we care about comes from Bayes' theorem, this means we can ignore the common scaling factor and simply use \eqref{eq:bayes2}: we only need to evaluate the prior and the likelihood, and the latter only requires that we evaluate the observation model $\observe(\state)$ of each candidate sample point. The Metropolis-Hastings method \cite{Chib1995} is the basic workhorse method in this class and is very simple to implement. If we also have access to the gradient of the observation operator, then we can use a hybrid (or Hamiltonian) Monte Carlo \cite{bishop2006pattern} to accelerate the sampling process. And importantly, if we suspect that there are multiple high-probability regions (i.e., multiple solutions of our inverse problem), then we can use the Method of Parallel Tempering \cite{Sambridge2014}, which uses multiple chains to more thoroughly explore the state space to identify these regions.

\subsubsection{Example: static estimation of point vortices from pressure}
\label{sec:vortex-static}

For a fluid dynamics demonstration of Bayesian inference with a nonlinear observation operator, we will consider a simple but illustrative example: using a small number of pressure sensors to estimate the instantaneous position and strength of one or more point vortices in an otherwise quiescent unbounded two-dimensional fluid domain. This example is covered in more detail by Eldredge and Le Provost \cite{eldredge2024bayesian}. The state vector $\state$ comprises $\statedim = 3\nv$ entries: the $(x_\vjdex,y_\vjdex)$ coordinates and strength $\Gamma_\vjdex$ of each of the $\nv$ point vortices. (We will assume we know $\nv$.) The problem is nondimensionalized by the fluid density and reference strength $\Gamma_{\text{ref}}$ and length $L_{\text{ref}}$. The observation vector $\meas$ consists of the $\measdim$ pressure measurements (relative to ambient pressure), $\Delta\press_\alpha$, $\alpha = 1,\ldots,\measdim$, at $\measdim$ sensor locations. The observation operator $\observe$ in this example is obtained analytically from the Bernoulli equation; it depends nonlinearly on all state components, quadratically on the strengths and inverse algebraically on the positions. Importantly, a pair of vortices, if comparably close to each other as they are to the sensors, generates a coupled component of the pressure that depends on the product of their strengths. The true measurement vector $\truemeas$ in this example is synthetically generated from the observation operator applied to the true state, $\truestate$, and adding Gaussian noise.

\begin{figure}
    \centering
    \includegraphics[width=0.48\linewidth]{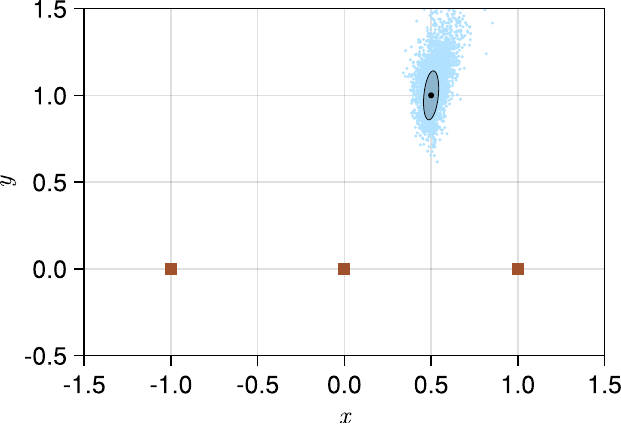}\includegraphics[width=0.48\linewidth]{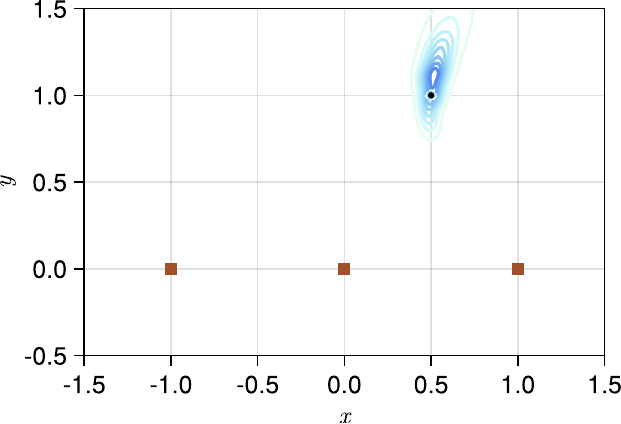}
    \caption{\newstuff{One point vortex estimated from three pressure sensors (brown squares). (Left) MCMC samples (light blue) in position plane, with true state (black dot) and covariance ellipse; (right) expected vorticity. Adapted with permission from Eldredge and Le Provost (J. Fluid Mech., 2024) \cite{eldredge2024bayesian}, copyright 2024 Jeff D. Eldredge.}}
    \label{fig:1vortex-3sensors}
\end{figure}

One of the inherent challenges of this vortex estimation problem is that there are often multiple likely states for a given observation vector. For example, consider three pressure sensors arranged in a straight line to estimate a single vortex, as depicted in Figure~\ref{fig:1vortex-3sensors}. With $\measdim = \statedim = 3$, there is ostensibly enough information in the pressure sensors to infer the vortex state. However, a vortex on either side of and at the same distance from this line of sensors would produce the same pressure measurements, as would a vortex of either sign of strength. Thus, there are four equally likely candidate solutions for any observation. Once again, the prior distribution becomes helpful. Here, we can use the prior to limit our estimation to a particular range of states based on prior knowledge, even if we know little else about the true state. For example, we can use a uniform distribution over vortices that lie to one side of the sensors and have positive strength. When using MCMC, such a prior prevents a chain entry from lying outside the acceptable range of states. 

Figure~\ref{fig:1vortex-3sensors} depicts the MCMC samples in the position plane and the resulting expectation of vorticity using a true observation from three noisy pressure sensors (with noise standard deviation of $5\times 10^{-4}$). The estimator is successful in estimating the location and strength of the vortex, though it has some uncertainty about the vertical position of the vortex. This uncertainty can also be visualized with the posterior covariance ellipse shown with the samples. This covariance was obtained analytically in this case by computing \eqref{eq:kalmanupdate-covar-linear} with $\observemat \approx \nabla\observe(\truestate)$, where $\truestate$ is the true state, and infinite prior covariance. Computing this covariance analytically obviously assumes that we know the true state, but it is useful at least for illuminating the information available in the observation operator and its gradient in a nonlinear problem. For example, it can be used to show that the maximum uncertainty of estimating a vortex grows quickly as the distance between the vortex and the sensors increases \cite{eldredge2024bayesian}. 

\begin{figure}
    \centering
    \includegraphics[width=0.48\linewidth]{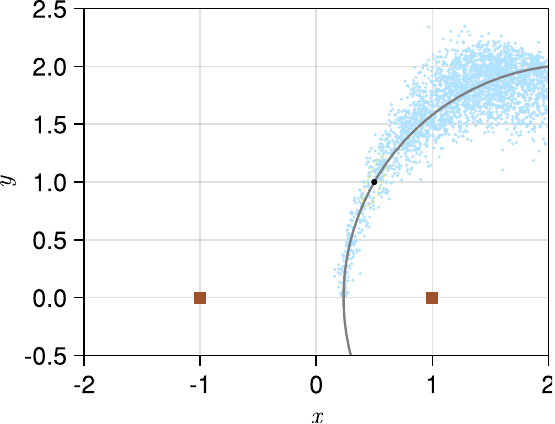}
    \caption{\newstuff{MCMC samples in the position plane of a single point vortex estimated from two pressure sensors (brown squares). Circle is the manifold of possible states, and black dot is the true state. Adapted with permission from Eldredge and Le Provost (J. Fluid Mech., 2024) \cite{eldredge2024bayesian}, copyright 2024 Jeff D. Eldredge.}}
    \label{fig:1vortex-2sensors}
\end{figure}

In the previous case, we had enough sensors to infer the vortex's characteristics. With only two sensors, we likely lack sufficient data to uniquely determine the vortex. Indeed, there is a manifold of equally-plausible states, and it can be shown that this manifold takes the form of a helix whose axis is parallel to the $\Gamma$ axis. In Figure~\ref{fig:1vortex-2sensors}, the MCMC samples are shown in the vortex position plane to be distributed about the projection of this manifold, a circle. We have no further knowledge to estimate the expected state. Another way to interpret this behavior is through $\observemat \approx \nabla\observe(\truestate)$. Similar to the manometer example, this matrix is rank deficient, in this case along a direction tangent to the manifold at $\truestate$. If the vortex state is perturbed along this manifold, the two sensors will not detect any change. In the manometer example, we simply fell back on the prior to estimate the uncertainty along the rank-deficient direction. Unfortunately, for the vortex, the prior covariance is infinitely large, so the posterior uncertainty remains infinite along the rank-deficient direction. We need more information about the vortex (via a better prior) or another sensor.

With the number of sensors at least as large as the number of states, $\measdim \ge \statedim$, we generally avoid issues of non-uniqueness in estimating a single point vortex. Suppose the vortex is not a point vortex, but has a spatially-distributed vortex core. Can we also estimate the vortex radius if we simply add a fourth pressure sensor? Unfortunately, estimating the size of a vortex from pressure data is problematic for two reasons \cite{eldredge2024bayesian}. First, the effect of a vortex's size on its pressure distribution is only appreciable if the sensors lie within the radius of the vortex core: the uncertainty grows like the inverse cube of the radius when the sensors are outside of this core. Second, even when the sensors lie inside the core, they have trouble distinguishing the states from one another, so the uncertainty remains large. Note that, as long as we do not seek to estimate the vortex size, the first behavior is actually useful: we can reliably estimate the position and strength of a real vortex with an observation model based on an idealized point vortex, as long as the real vortex is not very large in extent.

 \begin{figure}
    \centering
    \includegraphics[width=0.48\linewidth]{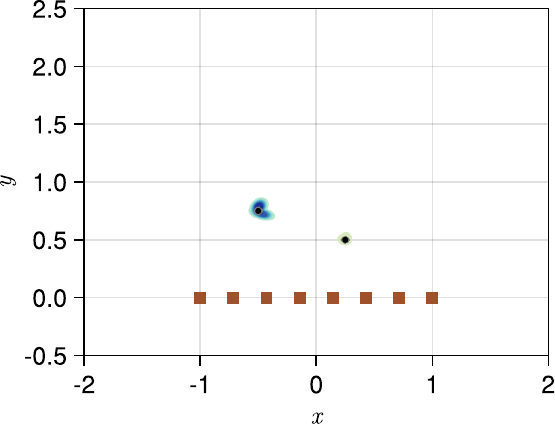}\includegraphics[width=0.48\linewidth]{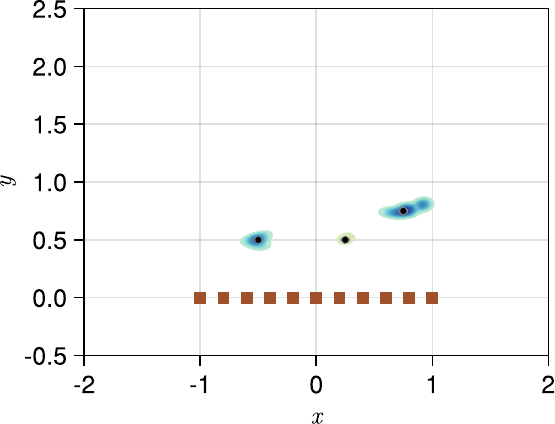}
    \caption{\newstuff{Expected vorticity of two point vortices estimated from 8 sensors (left) and three point vortices from 11 sensors (right). Adapted with permission from Eldredge and Le Provost (J. Fluid Mech., 2024) \cite{eldredge2024bayesian}, copyright 2024 Jeff D. Eldredge.}}
    \label{fig:multiple-vortices-static}
\end{figure}

With a sufficient number of pressure sensors, we can also estimate the characteristics of multiple vortices, as Figure~\ref{fig:multiple-vortices-static} illustrates. However, there are typically some isolated configurations of these vortices in which rank deficiency arises even when $\measdim = \statedim$. We avoid all rank-deficiency issues by simply using at least one additional sensor, $\measdim \ge \statedim + 1$. Even then, new challenges arise when estimating multiple vortices. In particular, a relabeling symmetry emerges: re-ordering the vortices in the state vector technically creates a new state, but obviously does not affect the pressure. Once again, this symmetry can be mitigated by restricting the prior, in this case to regions of the state space in which the vortices are ordered in some specific fashion, e.g., by their $x$ coordinate. Alternatively, we could have restricted the vortices in the prior to non-overlapping regions in the state space, provided the true vortices were covered by these regions. 

The aforementioned symmetry with respect to sign of strength is still present for multiple vortices. However, this quadratic dependence of pressure on strength also helps us, for the terms that couple vortices' effect on the pressure enable us to sense differences if only one vortex's sign is switched. Thus, the set of vortices can be estimated to within an overall sign, which can be established in the prior.


\subsection{Comments on static estimation}
\label{sec:gramians}

The examples in this section involving static estimation lead us to some key conclusions about uncertainty quantification and estimation of flows.

\paragraph{Spectral decomposition of the observation operator is important.} Singular value decomposition of the observation matrix \eqref{eq:Hdecompose} allows us to distinguish the informative and noninformative subspaces in the state and measurement spaces; the noninformative (non-observable) part is associated with the singular values that are either very small or exactly zero. This partition allows us to determine the state directions in which we are most uncertain. In cases in which we have more sensors than states, it allows us to determine the directions in the $\measdim$-dimensional measurement space that are most informative to the state estimate. And finally, it can be used to develop a rank-reduced form of the estimation algorithm to avoid potential corruption by the noninformative subspaces.

In estimation problems with nonlinear observation models, we would like to exploit the same benefits of the spectral decomposition of $\observemat$ to break down the estimation process. The obvious candidate for this decomposition is the Jacobian $\nabla\observe$, based on a Taylor expansion of the observation model. In probability theory, the relevant tool is the Fisher information matrix, which provides a general measure of the amount of information about the state $\state$ that is available in the observation $\meas$. For a Gaussian distribution of $\meas$, the Fisher information is given by $\nabla\observe(\state)^T \noisecovar^{-1}\nabla\observe(\state)$. But at which state do we evaluate this matrix? In the vortex example, we evaluated it at the true state, but this is hardly a useful strategy since we generally do not know this state. Cui and Zahm \cite{cui2021data} propose that we take the expected value of this Fisher information over the prior, resulting in the {\em state-space Gramian},
\begin{equation}
\label{eq:xgram}
    C_\state = \int \left(\noisecovar^{-1/2}\nabla\observe(\state)\statecovar^{1/2}\right)^T \left(\noisecovar^{-1/2}\nabla\observe(\state)\statecovar^{1/2}\right)\probdist_0(\state)\,\mathrm{d}\state \in \Reals^{\statedim\times\statedim}.
\end{equation}
In the case of a linear observation model, decomposed as in \eqref{eq:Hdecompose}, this Gramian reduces to $C_\state = VS^TSV^T$. In other words, the spectral decomposition of $\observemat$ automatically provides us with the eigendecomposition of $C_\state$, whose eigenvalues are in the diagonal matrix $\Lambda_\state = S^TS$ (and clearly non-negative) and whose eigenvectors are the columns of $V$. Spantini et al.~\cite{spantini2015optimal} showed that the eigenvectors with the largest eigenvalues indicate the state directions in which measurements are most informative relative to the prior---that is, the dominant directions in which the state will be updated by an observation. This suggests that, for a nonlinear observation operator, we can compute the eigenvectors $V$ and eigenvalues $\Lambda_\state$ of $C_\state$ defined by \eqref{eq:xgram} and reap the same benefits as for a linear operator. If we only have access to an ensemble of prior samples of the state, then \eqref{eq:xgram} can be approximated by a simple average over the ensemble, as shown in \eqref{eq:xgram-approx}.

What about the basis $U$ of the measurement space? Le Provost et al.~\cite{le2022low} suggest an analogous {\em measurement-space Gramian},
\begin{equation}
  \label{eq:ygram}
  C_\meas =  \int \left(\noisecovar^{-1/2}\nabla\observe(\state)\statecovar^{1/2}\right) \left(\noisecovar^{-1/2}\nabla\observe(\state)\statecovar^{1/2}\right)^T\probdist_0(\state)\,\mathrm{d}\state \in \Reals^{\measdim\times\measdim},
\end{equation}
which can also be approximated by averaging over an ensemble of prior samples (see \eqref{eq:ygram-approx}). This Gramian is not associated with a Fisher information matrix, but is inspired by the signal-to-noise ratio. Indeed, for a linear observation model, the decomposition of $\observemat$ leads to $C_\meas = USS^TU^T$, showing that the eigenvalues and eigenvectors of this Gramian matrix are $\Lambda_\meas = SS^T$ and $U$, respectively. As we saw in the manometer example, the dominant directions in $U$ place more weight on sensors with more signal relative to noise. By extension, the eigendecomposition of $C_\meas$ for a nonlinear operator gives us access to a basis for the measurement space with similarly useful properties. It needs to be stressed that, for cases with nonlinear observation operators, the eigenvalues $\Lambda_\meas$ and $\Lambda_\state$ will generally be different.

\paragraph{A good prior is very useful.} As we have seen, the prior distribution enables the estimator to make up for shortfalls in the informativeness of the measurement data or to select the best candidate from multiple likely solutions. In the sections that follow, we will target the prior for improvement in two ways. First, in \sect~\ref{sec:sequential_inference}, we will exploit sequences of observations over time, so that the prior is the posterior from the previous step, first advanced by a forecast of the state's evolution. Then, in \sect~\ref{sec:neural_network}, we will discuss the use of representations that assemble the flow state from a lower-dimensional manifold learned from data, so that the prior can restrict itself to states that are more likely to arise, based on previous training.


\section{Sequential estimation of flows}\label{sec:sequential_inference}


When the state evolves in time, then we seek an estimation approach that leverages two things: the information we already have about the state from past measurements, and any model at our disposal for predicting the state's advancement. For the latter, we will suppose we have a forecast model \eqref{eq:forecast} that expresses the dynamics of the state, but with process noise $\pnoise$ drawn from some distribution,
\begin{equation}
\label{eq:noisyforecast}
    \state_\timeindex = \forecast_\timeindex(\state_{\timeindex-1}) + \pnoise_\timeindex,
\end{equation}
where $\timeindex$ represents the time level. (We have omitted a possible dependence on a control input here, but it is straightforward to include it in everything that follows.) Note that this forecast only depends on the previous state and nothing earlier, indicating that this is a Markov process. For example, the Markovian aspect is met by the common forecast models in fluid dynamics, e.g., a time step of the Navier-Stokes or Euler equations or some reduced-order version of these equations. In this context, process noise physically represents random fast-varying dynamics that are not correlated in time. Probabilistically, this noisy forecast can be expressed through a {\em transition probability},
\begin{equation}
\label{eq:transition}
    \probdist(\state_\timeindex|\state_{\timeindex-1}),
\end{equation}
similar to the way the observation model is expressed as a likelihood. If we assume that $\pnoise_\timeindex$ is drawn from a Gaussian distribution $\normaldist(0,\pnoisecovar)$, then this transition probability is given by a Gaussian whose mean is the forecast operator, $\normaldist(\state_\timeindex|\forecast_\timeindex(\state_{\timeindex-1}),\pnoisecovar)$.

How do we use this transition probability? Suppose we start with a distribution $\probdist(\state_0|\meas_0)$ for the initial state $\state_0$ conditioned on one observation, $\meas_0$, obtained via the static inference we described in the previous section. To advance this to a distribution over all new states, $\state_1$, we marginalize over all possible initial states,
\begin{equation}
\label{eq:chapkol}
    \probdist(\state_1|\meas_0) = \int \probdist(\state_1|\state_0) \probdist(\state_0|\meas_0)\,\mathrm{d}\state_0.
\end{equation}
This distribution over $\state_1$ then serves as a prior for folding in a new observation $\meas_1$ with Bayes' theorem \eqref{eq:bayes1}, as we did before, to obtain a posterior distribution,
\begin{equation}
\label{eq:bayes-sequential}
    \probdist(\state_1|\meas_1,\meas_0) = \frac{\like(\meas_1|\state_1) \probdist(\state_1|\meas_0)}{\probdist(\meas_1)}.
\end{equation}
At the end of this two-step process, we have a distribution of states conditioned on all previous measurements, and clearly this process can be repeated indefinitely. \newstuff{These equations constitute an exact general framework for advancing the distribution, usually called recursive Bayesian estimation (or a Bayes filter) \cite{arulampalam2002tutorial}.} In the context of filtering, the first step is called the prediction, or {\em forecast}; the second is called the update, or {\em analysis}.

In the discussion that follows, we will generally assume that the forecast and observation operators are derived from physics. However, they can also be learned from data, and we will provide a more thorough discussion of estimation and uncertainty with learned operators in \sect~\ref{sec:neural_network}. However, the data can also be used to learn the transition probabilities and likelihood directly, and we highlight here an aerodynamics example of this approach in the recent cluster-based Bayesian model by Kaiser et al.~\cite{kaiser2024cluster}. In that work, the goal was to estimate the force on a delta wing from the measurement of a few surface pressure sensors. They assembled a large data set from experimental measurements of surface pressure and force on a delta wing at different angles of attack and undergoing various maneuvers. The force and pressure measurements form a joint state space, in which they first identified clusters, and from the data sequences, inferred the transition probability from each cluster to another. With these state transitions, and with a Gaussian model for the likelihood of a pressure observation belonging to a certain cluster, they used Bayes' theorem to estimate the posterior distribution of force at each time step.

\subsection{Linear operators, Gaussian noise}

For linear models of forecast $\forecastmat_\timeindex$ and observation $\observemat_\timeindex$ and Gaussian process and measurement noise, $\pnoise\sim \normaldist(0,\pnoisecovar)$ and $\noise\sim\normaldist(0,\noisecovar)$, the state distributions themselves remain Gaussian, and the filtering equations reduce exactly to equations for the state mean and covariance for both the forecast and the analysis, just as they did for the Bayesian inference problem on its own. This is the classical Kalman filter \cite{kalman1960new}, with a forecast step
\begin{align}
\label{eq:kalman-forecast}
    \mean{\timeindex|\timeindex-1} &= \forecastmat_\timeindex \mean{\timeindex-1}, \\
    \covar{\timeindex|\timeindex-1} &= \forecastmat_\timeindex \covar{\timeindex-1}\forecastmat_\timeindex^T + \pnoisecovar,
\end{align}
and an analysis step, based on new measurement $\truemeas_\timeindex$,
\begin{align}
\label{eq:kalman-analysis}
    \mean{\timeindex} &= \mean{\timeindex|\timeindex-1} + \kalman(\truemeas_\timeindex-\observemat_\timeindex \mean{\timeindex|\timeindex-1}), \\
    \covar{\timeindex} &= (\ident_\statedim - \kalman \observemat_\timeindex) \covar{\timeindex|\timeindex-1},
\end{align}
where $\kalman$ is given as before by \eqref{eq:kalman2}, using the forecast covariance $\covar{\timeindex|\timeindex-1}$ as the prior. A key point that we made in the discussion about $\kalman$ in the static inference discussion bears repeating. If this prior covariance becomes very small, then $\kalman$ tends to zero. This may seem like a good thing, because it seems to imply we are becoming more confident in our prediction, but it is actually catastrophic for data assimilation: the filter stops using measurements to update the state. Thus, we aim to preserve a certain level of distrust in our forecast if we want to keep using measurement data.

An example of a use of a linear Kalman filter for unsteady aerodynamics estimation is provided by An et al.~\cite{an2021lift}. In that work, the objective was to use sparse surface pressure measurements to estimate the lift coefficient on a NACA 0009 airfoil in a steady free stream undergoing large-amplitude pitch oscillations. The state vector comprised the lift coefficient and the pressure sensor values; including the measurements in the state ensured that the observation matrix simply picked off these measurement entries to predict the observation. A linear forecast operator for the lift coefficient was constructed from a modified Goman--Khrabrov model \cite{goman1994state} for the lift, combined with a least-squares regression for the lift from the pressure measurements, trained from a single case.

\subsection{Nonlinear operators}

For nonlinear forecast and/or observation operators, we are faced with the same challenge as in the static case, that the distributions arrived at in \eqref{eq:chapkol} and \eqref{eq:bayes-sequential} are not Gaussian, even if the prior and noise are. There are several options for how to proceed.

We could approximately solve the nonlinear filtering equations using MCMC, as we showed in the static inference example: this is the idea behind a particle filter. However, particle filters are expensive because \eqref{eq:chapkol} requires that we generate a set of forecast samples of $\state_1$ for every sample of $\state_0$ in the prior. For example, this would require advancing Navier-Stokes for a large number of initial conditions, each with a large number of instances of process noise, an intractable procedure.

Alternatively, if we assume that the distributions remain reasonably compact, then we can assume that they remain Gaussian. In that case, we could linearize each operator about the most recent mean state and use the standard Kalman filter \eqref{eq:kalman-forecast} and \eqref{eq:kalman-analysis}. In this approach, called the extended Kalman filter, the matrix operators $\forecastmat_\timeindex$ and $\observemat_\timeindex$ are replaced by Jacobians of the nonlinear operators evaluated at the respective means. Hemati et al.~\cite{hemati2014wake} used such an approach for estimating the parameters of line vortices that represented the wake of a lead aircraft in a formation, from pressure sensor data along the span of a trailing aircraft. These sensor measurements were interpreted with an observation model built from the Bernoulli equation and a lifting-line model of the trailing aircraft's own wake distribution. The extended Kalman filter may perform poorly when the operators are highly nonlinear, since a single Jacobian evaluated at the mean is insufficient for characterizing the uncertainty propagation. Instead, the unscented Kalman filter \cite{wan2000unscented} represents each state distribution with deterministically-chosen ``sigma points'', at which we evaluate the operators and then construct the necessary covariances to apply the Kalman steps.

For systems of large state dimension $\statedim$, as we frequently encounter in fluid dynamics, neither the extended nor the unscented Kalman filter is practical, since they both require explicit construction of covariance matrices that are $\statedim\times\statedim$. Some researchers have addressed this by working in a reduced-order basis for the flow field. Fukumori and Malanotte-Rizzoli \cite{fukumori1995approximate} proposed this approach in the context of oceanographic flows, and it was subsequently adapted by Suzuki \cite{suzuki2012reduced} to estimate shear flows by assimilating velocimetry data into a coarse direct numerical simulation (DNS). We will have more to say about reduced-order bases in \sect~\ref{sec:neural_network}.

Instead of computing these covariances explicitly, we could pursue something more like the particle filter, using an ensemble of states to approximate the distributions. But to avoid the computational drawbacks of the particle filter, we will assume that these distributions remain Gaussian, enabling us to draw upon some aspects of the Kalman filter. This is the foundation of the ensemble Kalman filter (EnKF), first proposed by Evensen \cite{evensen1994sequential} in the context of weather forecasting. The basic algorithm is as follows: For an ensemble of states $\{\statemember{\ensdex}\}_{\ensdex=1}^{\ensdim}$ drawn from some prior distribution, the forecast step is just a propagation of each ensemble member $\ensdex=1,\ldots,\ensdim$ through the forecast operator,
\begin{equation}
\label{eq:enkf-forecast}
    \statemember{\ensdex}_{\timeindex|\timeindex-1} = \forecast_\timeindex(\statemember{\ensdex}_{\timeindex-1}) + \pnoisemember{\ensdex}_\timeindex,
\end{equation}
and the analysis step is a Kalman update on each ensemble member, based on the innovation,
\begin{equation}
\label{eq:enkf-analysis}
    \statemember{\ensdex}_{\timeindex} = \statemember{\ensdex}_{\timeindex|\timeindex-1} + \kalman \left( \truemeas_\timeindex - \observe_\timeindex(\statemember{\ensdex}_{\timeindex|\timeindex-1}) - \noisemember{\ensdex}_\timeindex\right),
\end{equation}
where the construction of the Kalman gain $\kalman$ can be done in a few ways, most of which make use of the fact that the exact Kalman gain is $\kalman = \covar{\state\meas}\covar{\meas\meas}^{-1}$, as shown in \eqref{eq:kalman-app}. In particular, in the basic stochastic EnKF (sEnKF) \cite{evensen1994sequential}, the Kalman gain is simply computed using ensemble approximations of these covariances, $\kalman = \approxcovar{\state\meas}\approxcovar{\meas\meas}^{-1}$, where the approximations are given by \eqref{eq:ensxycov} and \eqref{eq:ensyycov}, using the ensemble that emerges from the forecast step. The advantage of the EnKF for high-dimensional problems (with ensembles smaller than the state dimension, $\ensdim < \statedim$) is that the computational costs of the forecast and analysis steps scale much better than for particle filters, since they simply involve $\ensdim$ evaluations of the forecast and observation operators, can be implemented using only existing codes for these operators, and require less storage than extended and unscented Kalman filters.

The sEnKF unfortunately exhibits a few undesirable behaviors for small- to modest-sized ensembles, where the computational advantages of the EnKF can best be realized. Much of this degraded performance can be attributed to two issues associated with a finite-sized ensemble: underestimation of the random process error in the forecast and spurious correlations in the ensemble covariances. The first of these issues is a common outcome of a deterministic forecast model (in which the process noise in \eqref{eq:enkf-forecast} is not naturally present) and causes a phenomenon called covariance collapse: the ensemble members all converge to the same value after several time steps, so the covariance matrices \eqref{eq:ensxxcov} and \eqref{eq:ensxycov} (and thus, $\kalman$) tend toward zero, preventing further updates of the ensemble from future measurements. It has become common practice to address this with a simple trick called ``covariance inflation'', which comes in two forms \cite{asch2016data}: a multiplicative form, in which each ensemble member's anomaly (i.e., its deviation from the mean) is multiplied by a factor $\beta$ just larger than 1; and an additive form, in which a random process noise $\pnoisemember{\ensdex} \sim \normaldist(0,\pnoisecovar)$ is added to each ensemble member. The choices for $\beta$ and $\pnoisecovar$ need not be tied to anything physical; they simply ensure a minimum level of uncertainty so that the ensemble remains responsive to new measurements.

The sEnKF with inflation has been used successfully in several flow estimation investigations. Colburn et al.~\cite{colburn2011state} used pressure and skin friction measurements at the wall of an incompressible turbulent channel flow to estimate the mean velocity field in the channel. Da Silva and Colonius \cite{da2018ensemble} used it to estimate the evolving velocity field of a pair of low Reynolds number flows: past a circular cylinder, using velocity sensors in the wake; and past a NACA 0009 airfoil at 30 degrees angle of attack, using surface pressure sensors. In both cases, they used direct numerical simulation of the incompressible Navier--Stokes equations and associated pressure Poisson equation to provide the forecast and observation operators. Truth measurements were provided by the same simulations, but with noise added to the observations. In the work of Darakananda et al.~\cite{darakjdeprf18}, surface pressure measurements were obtained (with noise added) from direct numerical simulation (DNS) of the Navier--Stokes equations of a flat-plate airfoil at 20 and 60 degrees angle of attack and Reynolds number 500, in some cases subjected to large impulsive flow disturbances. They interpreted these measurements with an inviscid point vortex model of the same geometry. An ensemble of 50 members was propagated with this model via the sEnKF with inflation. In that work, the estimator's knowledge of the flow disturbance was only obtained from the surface pressure measurements of the DNS, and these were shown to successfully update the vortices' positions and strengths in the analysis step to mimic the disturbance's convective influence. 

\begin{figure}[t]
    \centering
    \includegraphics[width=0.8\linewidth]{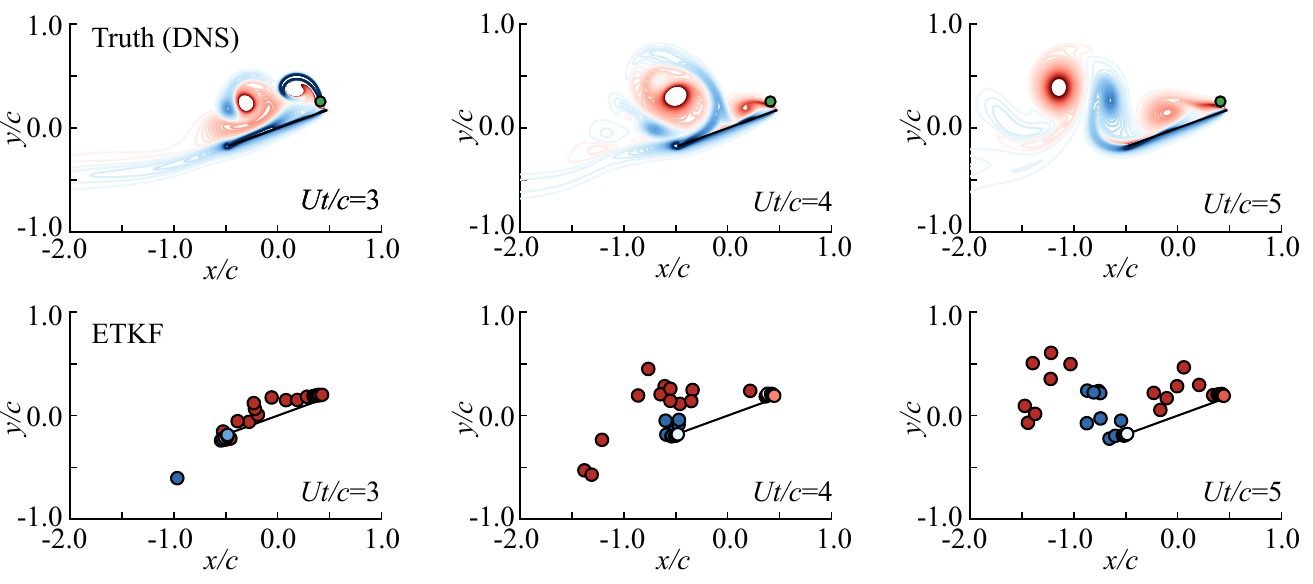}
    \includegraphics[width=0.6\linewidth]{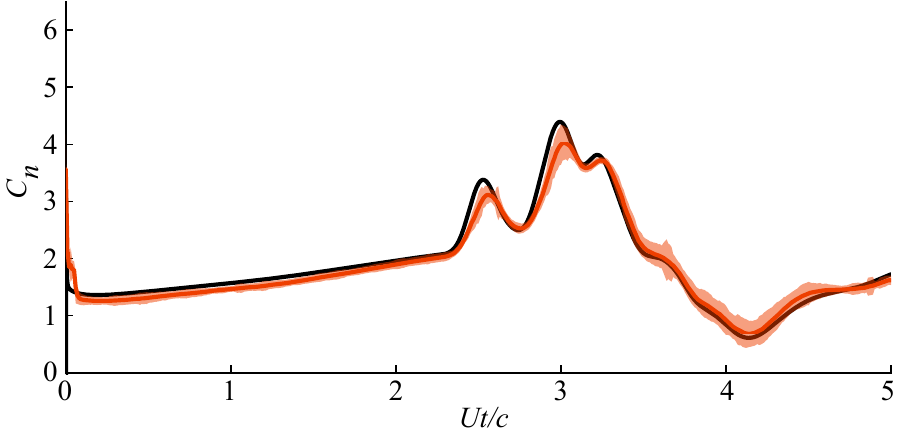}
    \caption{\newstuff{Comparison of true flow (DNS, Re = 500) and point-vortex estimate, from DNS surface pressure. Adapted with permission from Le Provost and Eldredge (Phys. Rev. Fluids, 2021) \cite{le2021ensemble}, copyright 2021 American Physical Society.}}
    \label{fig:vortex-estimation-Re500}
\end{figure}

The relatively large ensemble size needed in \cite{darakjdeprf18} reflects the second issue with sEnKF, due to spurious correlations. This issue can be addressed in a few ways, depending on the origin. For example, the measurement noise $\noisemember{\ensdex}$ and state $\statemember{\ensdex}$, though ideally uncorrelated, are not uncorrelated in a finite ensemble. A particular consequence is that the posterior and prior covariances do not obey the ideal relationship in \eqref{eq:kalmanupdate-covar-linear}, $\covar{\state|\truemeas} = (\ident_\statedim - \kalman \observemat)\statecovar$, that expresses the analysis step's reduction of our uncertainty. Thus, many of the pernicious effects of noise--state correlation are mitigated with the ensemble transform Kalman filter (ETKF), developed by Bishop et al.~\cite{bishop2001adaptive}, in which the Kalman update is re-derived to exactly reproduce the ideal relationship for any ensemble size. Le Provost and Eldredge \cite{le2021ensemble} showed that the ETKF can significantly improve the performance of the sEnKF for aerodynamic flows similar to those considered in \cite{darakjdeprf18}. An example of those results is shown in Figure~\ref{fig:vortex-estimation-Re500}, which depicts, at three instants during the disturbance response, a comparison of the true vorticity from DNS with the ensemble mean vortex positions of the estimator. A comparison of estimated and true normal force is also shown.

The spurious correlations among states and between states and observations are more complicated to address because they are necessarily mixed in with true physical correlations. The technique of ``localization'' \cite{asch2016data} can often effectively address this by restricting covariance entries to those in which the state and the sensor are close in some geometric sense. However, many observations in fluid dynamics involve algebraic relationships over long distances, e.g., pressure and velocity, and localization can artificially truncate these relationships. Thus, we present a variant of the EnKF in the next section that provides an alternative approach to offset the challenges from spurious correlations.

\subsection{The low-rank EnKF} 

Before we present the low-rank EnKF (or LREnKF) \cite{le2022low}, it is important to note that the ensemble covariances \eqref{eq:ensxxcov}--\eqref{eq:ensyycov} are limited to rank $\ensdim-1$, as discussed below those equations in Appendix \ref{sec:appendix-ensemble}. Thus, if the number of sensors and the state dimension are both larger than the ensemble size, the state update \eqref{eq:enkf-analysis} in the sEnKF---by virtue of \eqref{eq:ensxycov}---is rank deficient, limited to a linear combination of the $\ensdim$ state anomalies $\statemember{\ensdex}-\approxmean{\state}$ in the prior ensemble (where $\approxmean{\state}$ is the ensemble mean \eqref{eq:ensxmean}). Without further treatment, the sEnKF weights this linear combination haphazardly, in a manner that can be corrupted by the aforementioned spurious correlations. It would be better if we could curate this linear combination to avoid these spurious influences.

How can we curate it? We saw in the case of the linear operator in \sect~\ref{sec:static_inference} that the columns of $V$ and $U$ form respective bases for the state and measurement spaces that, by the magnitude of the singular values, partition the informative from the noninformative subspaces of these spaces; for a nonlinear observation operator in an ensemble, we can make use of the eigenvectors and corresponding eigenvalues of the Gramians \eqref{eq:xgram-approx} and \eqref{eq:ygram-approx} for this purpose, as we discussed in \sect~\ref{sec:gramians}. It is important to stress that the ``noninformative'' subspace is not concretely defined and can include directions with eigenvalues that are relatively small compared to the dominant ones. In fact, the most pernicious aspects of the finite-sized ensemble typically lie in these smaller-eigenvalue directions, so we expect better behavior if we simply avoid them. This is the foundational idea of the LREnKF \cite{le2022low}. We truncate each basis at respective ranks $\staterank$ and $\measrank$, chosen so that the cumulative fraction of energy in the retained directions just exceeds some threshold, $\alpha$, i.e., $\sum_{j=1}^{\staterank} \lambda_{\state,j}/\sum_{j=1}^{\statedim}\lambda_{\state,j} \geq \alpha$ and $\sum_{j=1}^{\measrank} \lambda_{\meas,j}/\sum_{j=1}^{\measdim}\lambda_{\meas,j} \geq \alpha$, in which the eigenvalues are sorted in decreasing order. After choosing the reduced ranks, the Kalman gain is constructed via the spectrally-decomposed form \eqref{eq:kalman-svd-def}, but now with the reduced bases $V_{\staterank}$ and $U_{\measrank}$:
\begin{equation}
\label{eq:kalman-reduced}
    \kalman = \statecovar^{1/2}  V_{\staterank} \approxcovar{\statesvd\meassvd} \approxcovar{\meassvd\meassvd}^{-1} U_{\measrank}^T \noisecovar^{-1/2},
\end{equation}
where the covariances $\approxcovar{\statesvd\meassvd}$ and $\approxcovar{\meassvd\meassvd}^{-1}$ are computed with the ensemble expressed in the coordinates of the reduced bases (see \eqref{eq:enscovar-reduced} and the equations leading up to it). The rank-reduced Kalman gain \eqref{eq:kalman-reduced} assembles the state update from a linear combination of the reduced basis $V_{\staterank}$ of the state space rather than from the original ensemble of state anomalies, thus achieving our goal. We add that, for general nonlinear observation operators, the Kalman gain in the new basis $\approxcovar{\statesvd\meassvd} \approxcovar{\meassvd\meassvd}^{-1}$ is not diagonal, so the measurement modes are not paired with individual state modes, as in the linear case.

There are two complications of the LREnKF compared to the sEnKF. The LREnKF needs access to the Jacobian of the observation operator, ideally by analytical means or automatic differentiation, which is more intrusive than the basic sEnKF. Furthermore, one must compute eigenvectors and eigenvalues of the Gramian matrices, which are of size $\Reals^{\statedim\times\statedim}$ and $\Reals^{\measdim\times\measdim}$, respectively. For large-dimensional systems, the first of these can be computationally intensive. However, the method enables the use of very small ensembles (smaller than either $\statedim$ or $\measdim$) without exhibiting filter divergence, as occurs for the sEnKF in such ensemble sizes \cite{le2022low}. Thus, there is a trade-off in cost that makes these extra computations worthwhile.

\subsection{Example: sequential estimation of point vortices}
\label{sec:vortex-sequence}

\begin{figure}
    \centering
    \includegraphics[width=1.0\textwidth, trim=0 80 0 80, clip]{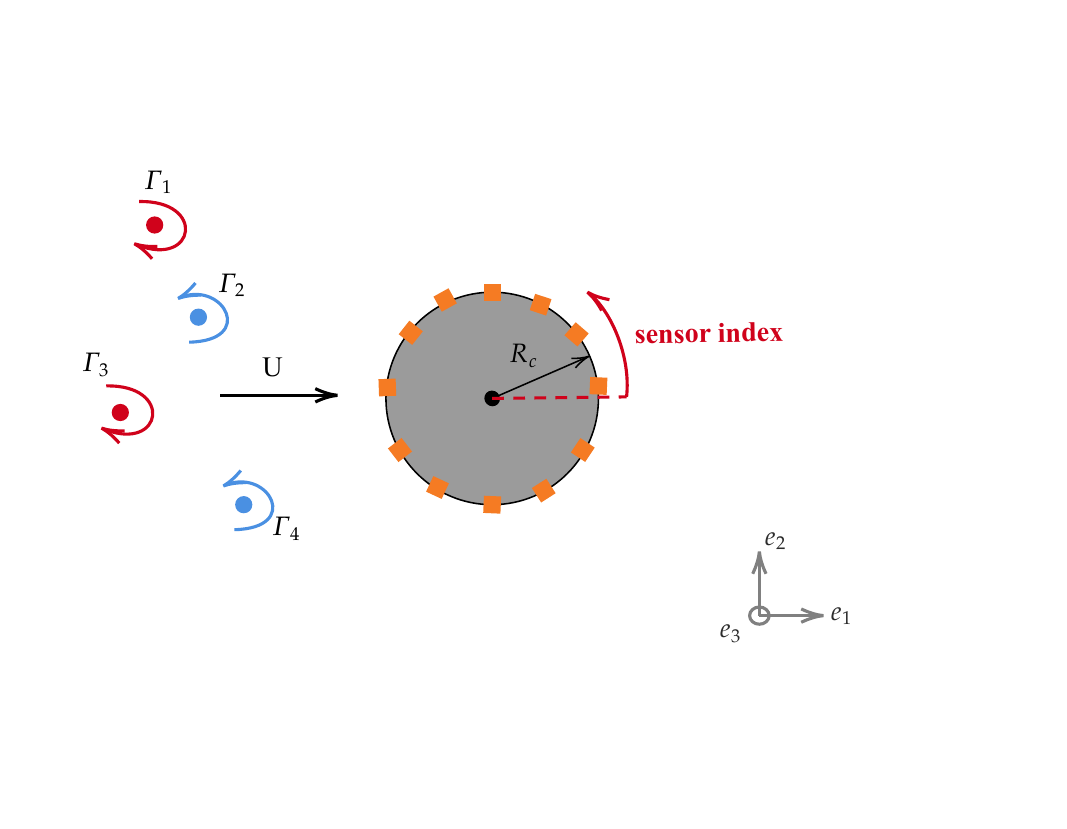}
    \caption{Schematic of the two-dimensional flow over a cylinder with large-scale structures represented by point vortices. Orange squares on the cylinder represent sensors.}
    \label{fig:cylinder_config}
\end{figure}

In \sect~\ref{sec:vortex-static} we discussed an example in which we used pressure measurements to infer the characteristics of one or more point vortices at a single instant. In that example, we assumed very little prior knowledge and used MCMC sampling to characterize the posterior probability. In the current example, we will also estimate point vortices from pressure measurements. However, now we will perform the estimation over an extended time interval as the flow evolves, taking in pressure measurements at a constant rate and using the LREnKF to assimilate these measurements into an ensemble that approximates the evolving probability. In this case, we will consider a true flow as depicted in Fig.~\ref{fig:cylinder_config}, consisting of point vortices self-advecting in a uniform flow of speed $U$ past a circular cylinder, and in which measurements are obtained from a uniformly-distributed array of pressure sensors on the cylinder surface. For the purposes of demonstration, we will assume the flow is inviscid, incompressible, and two-dimensional. We obviously recognize that this true flow is not representative of a realistic vortex encounter in a viscous flow. Our purpose here is merely to illuminate the general aspects of the observability of vortices.

As in the previous example, the state vector includes the positions and strengths of the vortices. The forecast operator comprises the time advancement of the positions of the point vortices by a forward Euler integration of the local velocity field for one time step; the observation is obtained from solution of the pressure Poisson equation. The velocity and pressure fields are obtained with a grid-based potential flow solver that uses an immersed boundary treatment of the body surface and lattice Green's function for the solution of Poisson equation; the details are described in \cite{beckers2022planar}. The grid spacing is $\Delta x/R_c = 0.02$ and the time step size is $U \Delta t / R_c=0.02$. The Jacobian of the observation operator, $\nabla\observe$, is evaluated using automatic differentiation of the aforementioned numerical operators. True pressure measurements $\truemeas_\timeindex$ are provided at every time step, and these are synthetically generated with the same forecast and observation operators as in the estimation, but starting from an initial state $\state_0$ unknown to the estimator and augmented with Gaussian measurement noise with standard deviation $\noisestddev=10^{-3}$.


\subsubsection{One true vortex and one-vortex estimator}
We begin by examining the case of a single vortex ($\statedim = 3$) approaching the cylinder. This simple configuration allows us to establish a framework for examining more complex interactions. The true vortex has unit strength, $\Gamma^*/UR_c = 1$, and is initially positioned upstream of the cylinder at $(x^*_0/R_c, y^*_0/R_c)=(-3,0)$. A set of $\ensdim=10$ ensemble members are drawn from an initial Gaussian distribution $\normaldist(\mean{0},\covar{0})$, where the mean is offset from the true state, $\mean{0} = \truestate_0 + \state_{\mathrm{off}}$, and the initial covariance matrix is diagonal, with variance $\sigma^2_x$ for both positions and $\sigma^2_\Gamma$ for the strength. The offset introduces an initial bias in the ensemble to represent a reasonable lack of initial knowledge of the vortex. We arbitrarily set this offset to $\state_{\mathrm{off}} = (x_{\mathrm{off}},y_{\mathrm{off}},\Gamma_{\mathrm{off}}) = (0.3,0.5,0.4)$ and the standard deviations to $\sigma_x=\sigma_{\Gamma}=0.5$. A total of $\measdim=40$ sensors are uniformly distributed along the surface of the cylinder. This dense sensor arrangement is chosen to capture the transient, localized response of the measurements to the vortices as they interact with the body. Additive inflation with variance of $10^{-8}$ and multiplicative coefficient of 1 are set during the inference to prevent ensemble collapse. The LREnKF is used for estimation with a rank threshold $\alpha=0.99$. We find that the measurement-space rank remains between 2 and 3 for the duration of the encounter, unsurprising considering that the untruncated rank is $\statedim=3$. 

\begin{figure}
    \centering
    \begin{overpic}[width=1\linewidth]{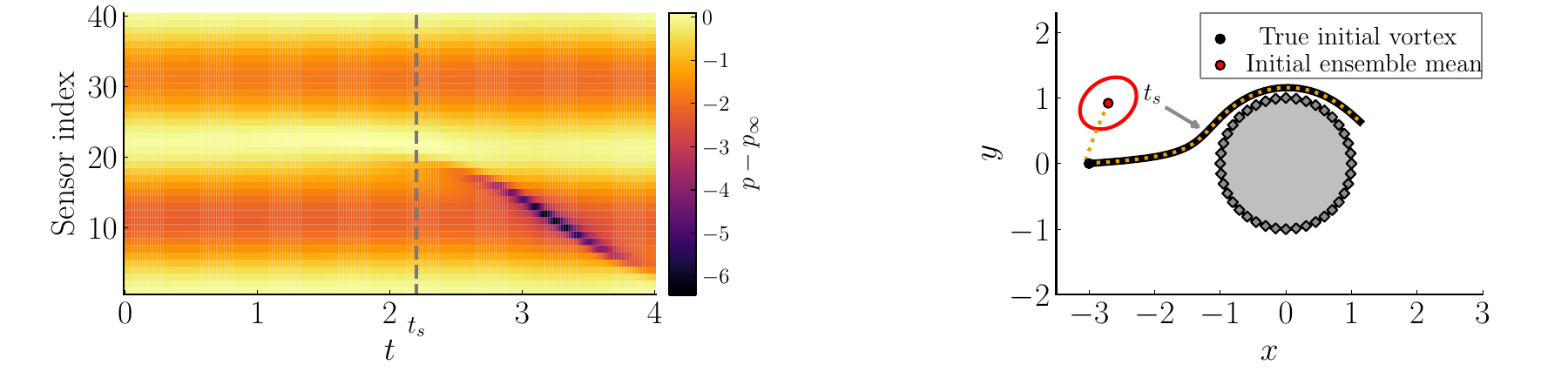}
    \put(-1,25){(a)}
    \put(55,25){(b)}
    \end{overpic}
    \caption{One true vortex and one-vortex estimator. (a) Surface pressure sensor histories (indexed as in Fig.~\ref{fig:cylinder_config}). (b) Vortex trajectories: true (black solid line) and estimated (yellow dotted line). Initial ensemble mean and covariance ellipse are shown. Surface sensors shown as gray squares.}
    \label{fig:1vortex_estimation}
\end{figure}

\begin{figure}
    \centering
    \begin{overpic}[width=1\linewidth]{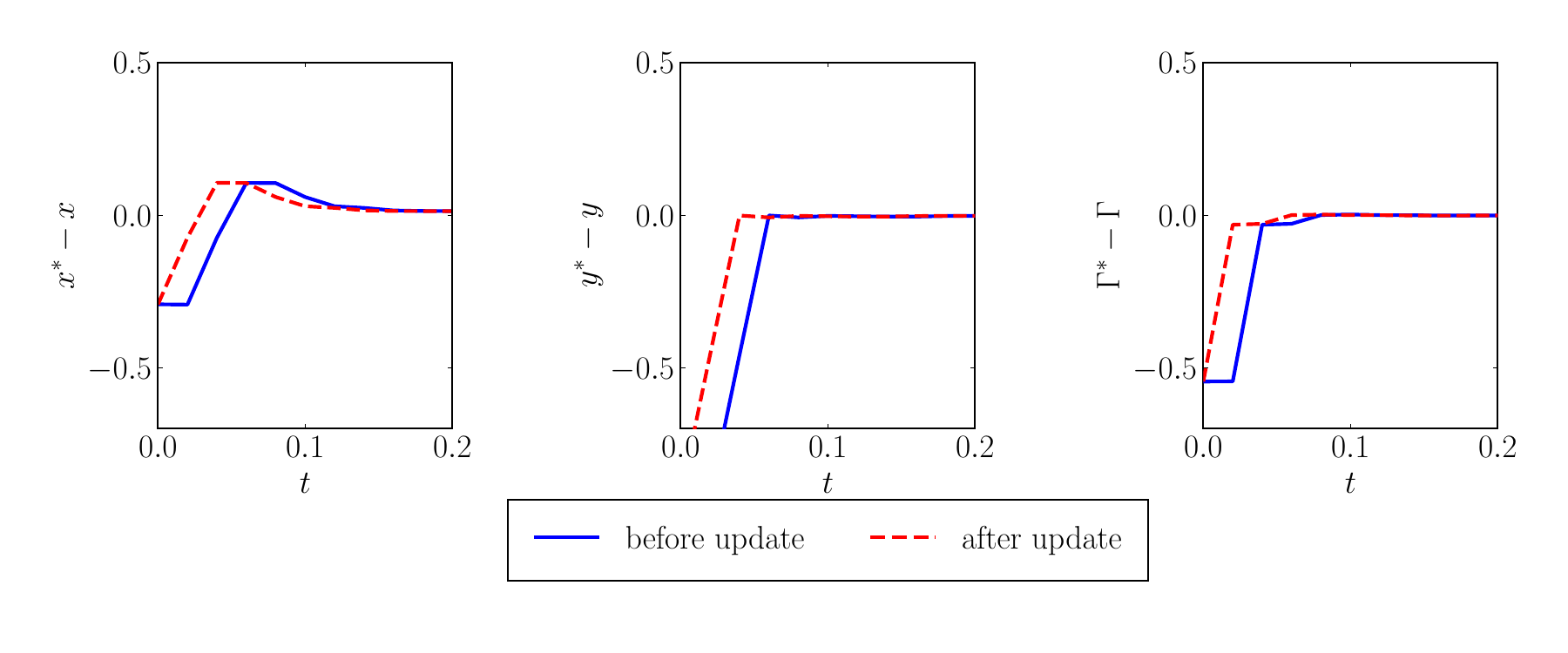}
    \put(-1,40){(a)}
    \put(34,40){(b)}
    \put(67,40){(c)}
    \end{overpic}
    \caption{One true vortex and one-vortex estimator. Estimation errors before and after the state update at early times for (a) $x$ coordinate, (b) $y$ coordinate, and (c) strength of the predicted vortex.}
    \label{fig:1vortex_error}
\end{figure}

Figure~\ref{fig:1vortex_estimation} illustrates the estimation results for the trajectory and strength of the vortex passing over a cylinder in uniform flow. The spatiotemporal plot in Fig.~\ref{fig:1vortex_estimation}(a) depicts the surface pressure recorded by all of the sensors over time. The surface pressure sensors positioned on the left side of the cylinder initially respond to the vortex's approach around $t=t_s$, when the vortex is closest to these sensors (see the location of the vortex at this moment in Fig.~\ref{fig:1vortex_estimation}(b)). As the vortex moves along the upper surface, sensors in its immediate vicinity become increasingly responsive. An ensemble of vortices is initially placed in the covariance ellipse illustrated in Fig.~\ref{fig:1vortex_estimation}(b), with the true initial vortex location shown for reference. The estimated trajectory of the vortex converges to the true trajectory after only three time steps and remains close to this trajectory with consistently low uncertainty set by the covariance inflation. The comparison of the ensemble mean's history with the true trajectory is depicted in Fig.~\ref{fig:1vortex_error} for the vortex coordinates and strength. The plots show the immediate effect of using measurements to update the state from its initial prior ensemble.

\begin{figure}
    \centering
    \includegraphics[width=1.0\textwidth]{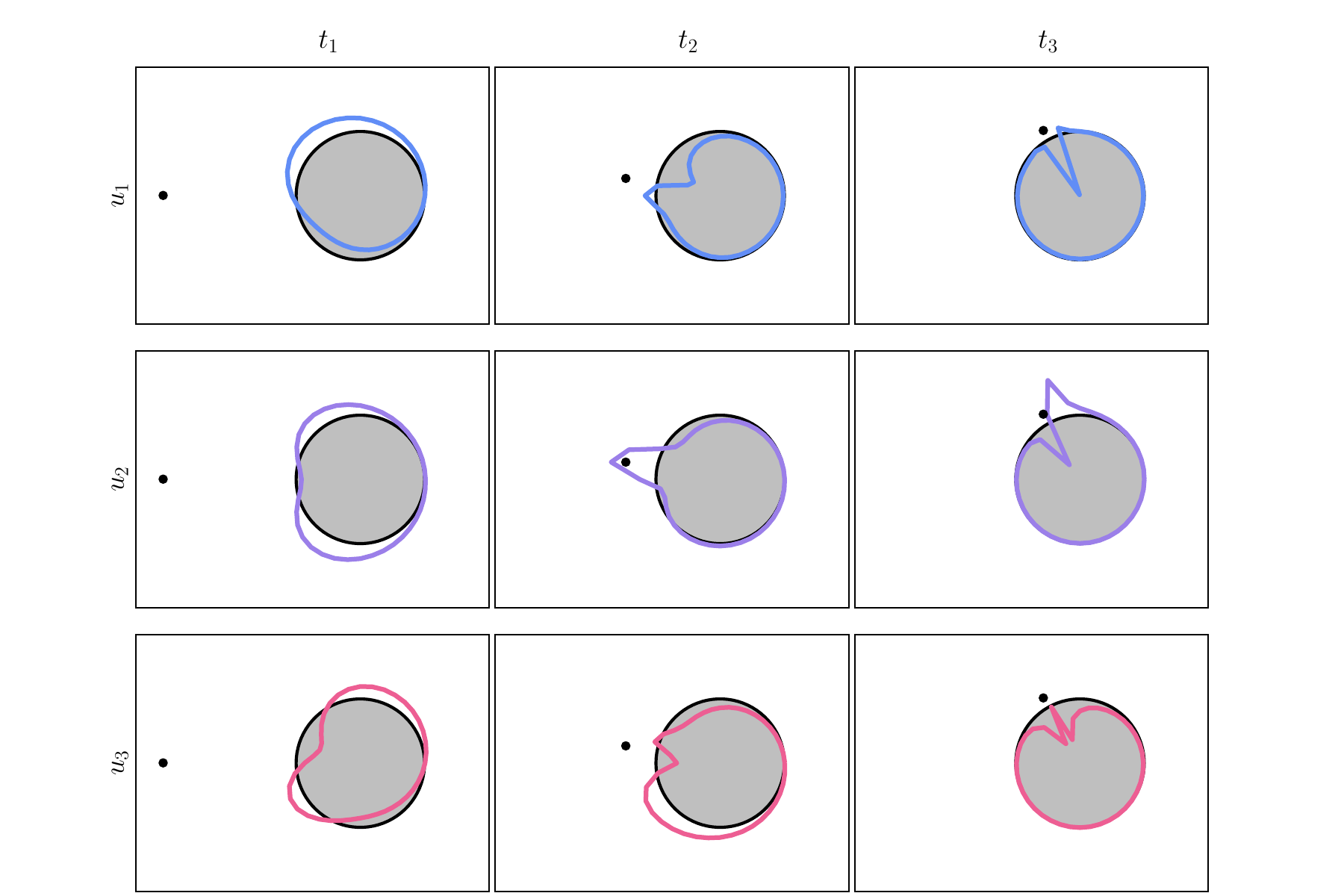}
    \caption{One true vortex and one-vortex estimator. The first three observation modes, $(u_1, u_2, u_3)$, of the measurement-space Gramian $C_\meas$ are depicted at three distinct time instants as a vortex (indicated by a black circle) moves past the cylinder.}
    \label{fig:1vortex_observation_modes}
\end{figure}

Figure~\ref{fig:1vortex_observation_modes} presents the first three eigenvectors of the measurement-space Gramian at three different time steps as the vortex passes by the cylindrical body. These modes demonstrate the dominant ways in which measurement data is aggregated to update the state estimate. The figure shows that when the vortex is far from the body, sensor readings contribute relatively broadly over the whole cylinder, suggesting that distance localization would be a poor strategy for improving the performance of the sEnKF. As the vortex nears the cylinder, measurements from sensors in its vicinity become more responsive to flow changes, and these dominant modes only gather information from a few sensors. Higher modes generally have more complex features and less spatial compactness. 

\subsubsection{Five true vortices and five-vortex estimator}
\label{sec:5vortex-5estimate}
As in the static inference case, the complexity of the problem increases significantly when multiple vortices are present in the flow. To demonstrate the capability of the LREnKF in handling (and illuminating) such complex estimation scenarios, we set up a case in which the true flow consists of five point vortices initially placed upstream of the cylinder in an arbitrary configuration. We use the LREnKF to estimate the trajectories and strengths of all five vortices from the measurements obtained from the array of surface pressure sensors. The true initial positions and strengths of the vortices are given by $(x^*_0/R_c,y^*_0/R_c,\Gamma/UR_c) = (-3,-0.8,-1)$, $(-2.8,-0.5,-1.1)$, $(-2.9,0,1.2)$, $(-2.7,0.4,1.3)$ and $(-2.8,0.8,1.4)$. An ensemble of $\ensdim=50$ members is used to estimate the $\statedim=15$ components of the state of the five vortices. The initial ensemble is drawn from a Gaussian prior with a mean that is offset from the true value, similar to the single vortex case. To offset the relabeling symmetry discussed in \sect~\ref{sec:sequential_inference}, the Gaussians associated with each vortex do not overlap with one another. The offsets are randomly drawn from uniform distributions $(x_{\text{offset}} \sim \mathcal{U}(0,0.4)$, $y_{\text{offset}} = 0$, and $\Gamma_{\text{offset}} \sim \mathcal{U}(-0.4,0.4))$, where $\mathcal{U}$ denotes a uniform distribution, and the initial prior's standard deviations for each vortex are $\sigma_x = 0.2$, $\sigma_y = 0.1$ and $\sigma_\Gamma=0.1$.

\begin{figure}
    \centering
    \begin{overpic}[width=1\linewidth]{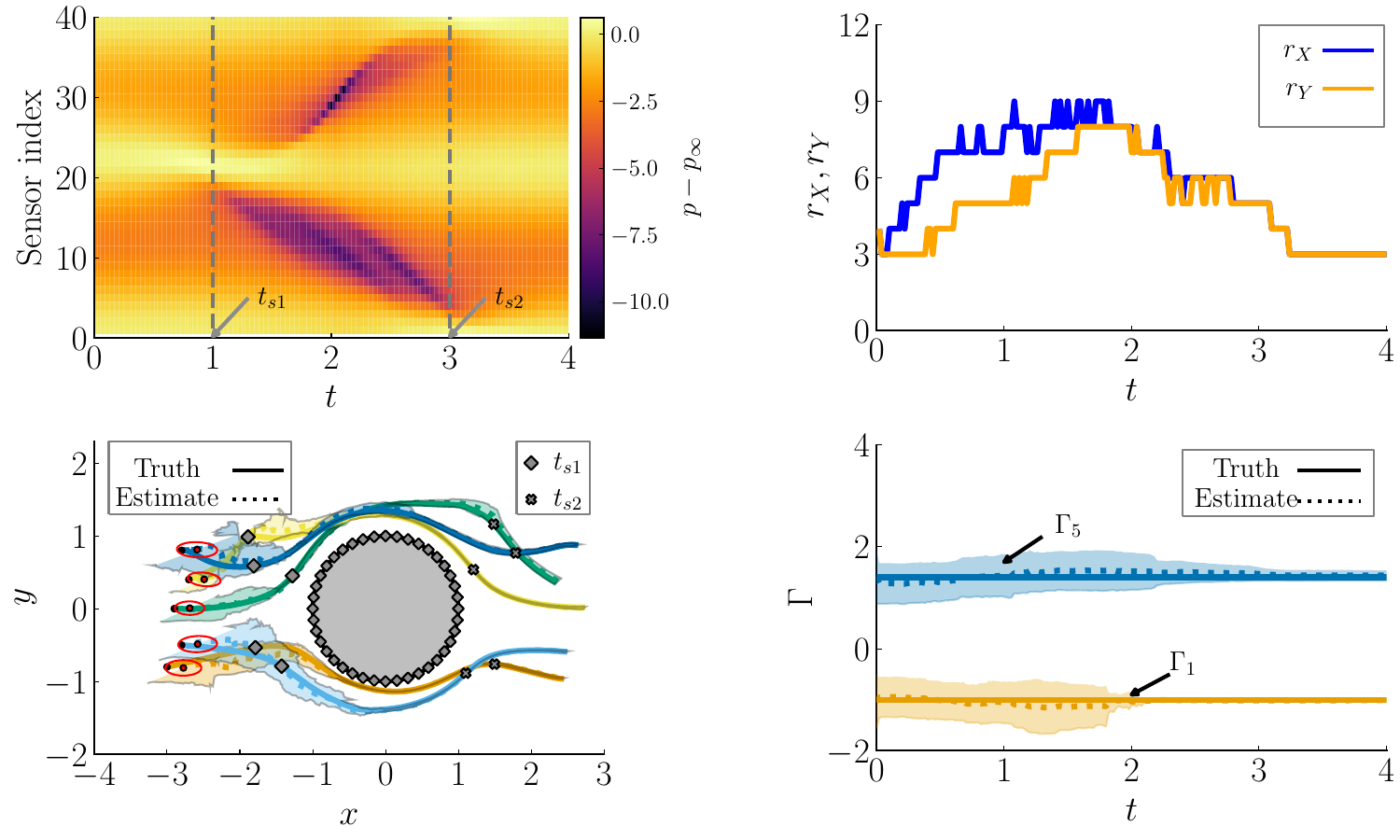}
    \put(-1,60){(a)}
    \put(53,60){(b)}
    \put(-1,28){(c)}
    \put(53,28){(d)}
    \end{overpic}
    \caption{Five true vortices and five-vortex estimator. (a) Surface pressure sensor histories. (b) Ranks of state and measurement Gramians. (c) Vortex trajectories, as in Fig.~\ref{fig:1vortex_estimation}. (d) True (solid lines) and predicted (dashed lines) circulations for vortices 1 and 5 (colors consistent with panel (c)), with $95\%$ confidence intervals.}
    \label{fig:5vortices_estimation}
\end{figure}

Figure \ref{fig:5vortices_estimation}(a) presents a spatiotemporal plot of the transient response of the surface pressure sensors during the vortex encounter. This plot clearly indicates that sensors at the front of the cylinder, and then those along the upper and lower sections, exhibit heightened responses as the vortices traverse the cylinder during the time interval $t \in [t_{s1},t_{s2}]$. For a detailed view of the true vortex positions at these specific instants, refer to Figure \ref{fig:5vortices_estimation}(c). This panel also shows the predicted trajectories of the five vortices, and Figure~\ref{fig:5vortices_estimation}(d) presents the predicted strengths of two of the vortices, all with associated $95\%$ confidence intervals. The initial ensembles are shown in panel (c) with their mean and covariance ellipses shown in red. The trajectories demonstrate the complex interactions among the vortices and with the body. Initially, with the vortices far from the body, the estimator predicts a high-uncertainty region around each trajectory, encompassing the true state. As the vortices approach the body, the estimator’s predictions generally become more accurate and uncertainty drops significantly. Uncertainty also tends to grow during intervals in which vortices come close to one another, and consequently, their forecast motions are most sensitive to small perturbations.

\begin{figure}
    \centering
    \begin{overpic}[width=1\linewidth]{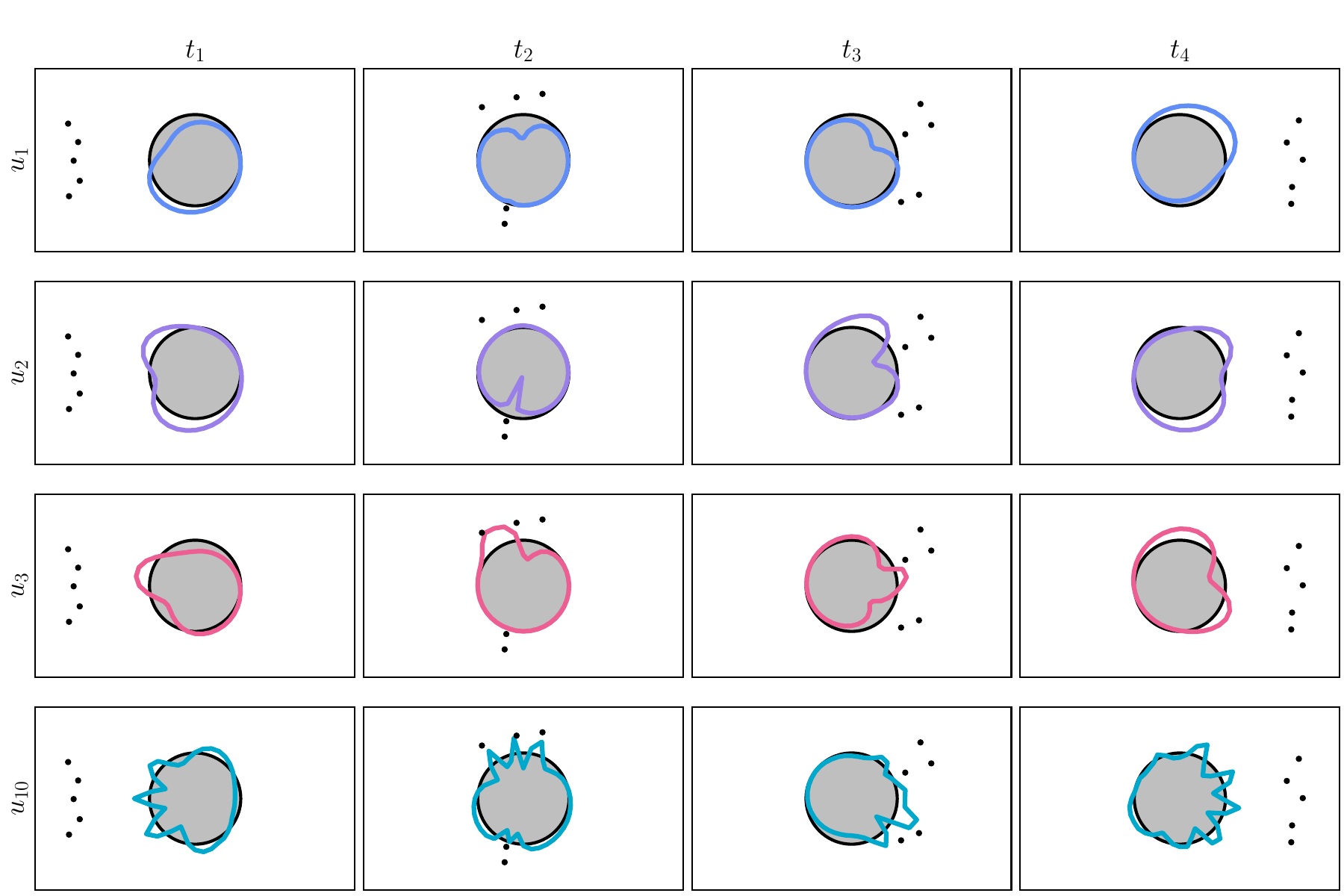}
    \put(-5,65){(a)}
    \end{overpic}\\
    \begin{overpic}[width=0.55\linewidth]{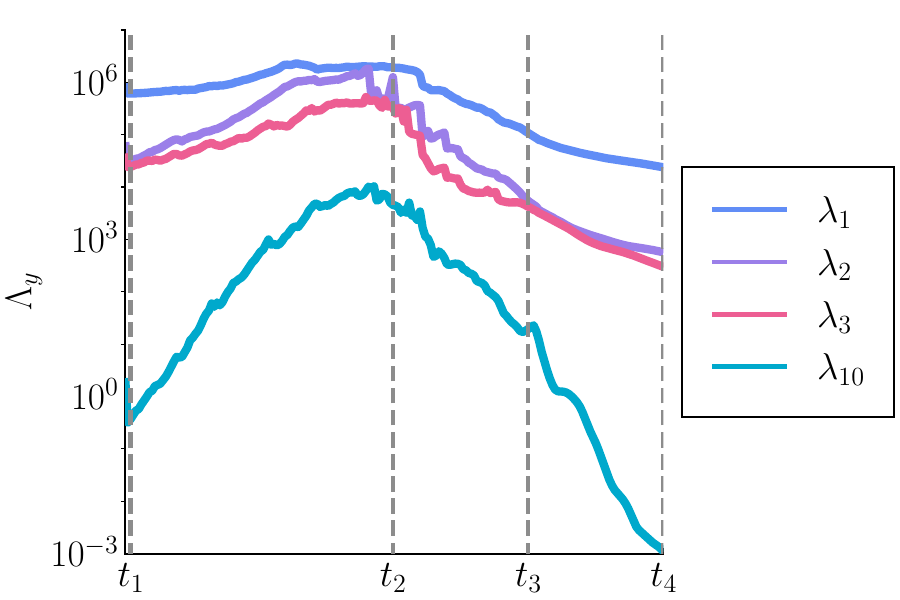}
    \put(-5,50){(b)}
    \end{overpic}
    \caption{Five true vortices and five-vortex estimator: (a) 1, 2, 3, and 10th eigenvectors of observation Gramian $C_\meas$, $(u_1, u_2, u_3, u_{10})$ are depicted at four distinct time instants as the vortices (indicated by black circles), (b) the evolution of 1, 2, 3, and 10th eigenvalues of the observation Gramian.}
    \label{fig:5vortices_observation_modes}
\end{figure}

The observed behavior can be better understood by viewing it through the reduced bases. Figure \ref{fig:5vortices_estimation}(b) depicts the evolution of the reduced ranks in the state and measurement spaces, chosen so that $99\%$ of the cumulative energy is retained in the eigenvalues of the respective reduced subspaces (i.e., $\alpha = 0.99$). For this specific setup, the measurement rank is observed to rise from an initial value of 3 to a maximum of 9 as the vortices approach the body, then decrease back to 3 as the vortices move away. In other words, during the vortex--cylinder encounter, only a small subset of directions in the measurement space contributes to updating the states. Similarly, the state update is only assembled from a modest number of state bases, as evident in the fact that the state rank starts at 3 and increases only to 9 when the vortices are closest to the cylinder.

These trends are remarkable. Based on our intuition from linear problems, with only 15 states we obviously only expect a comparable number of directions in the measurement space to be informative rather than the full $\measdim=40$. However, the results indicate that, when the vortices are far from the body, the information available in the sensors is limited to 3 modes that update 3 overall features of the vortex array. As the vortices move closer to the body, the sensors record more information about the vortices' individual characteristics and are able to provide a more thorough update of the state of the vortices. But this update's highest rank is still short of the full 15 dimensions, indicating that these vortices are never fully observable at any one instant. However, it is important to emphasize that the informative subspaces evolve over time, so that even if a particular state component is less responsive to sensors at one instant, it may become more responsive at later times, and vice versa. The prior conveys the uncertainty through these sequential updates, enabling precision in some state directions to be preserved even when those directions cannot be updated over some intervals. 

The manner in which the sensor information is gathered is apparent in the column vectors of $U_{\measrank}$, a few of which are depicted in Figure~\ref{fig:5vortices_observation_modes}(a). The dominant modes reflect a response of the sensors to clusters of vortices; this response is relatively diffuse over the full set of sensors when the clusters are far away from the sensors, and more localized when they are close. The higher modes engage a broader array of sensors and depict more individual responses to the vortices. However, as the eigenvalue trends in Figure~\ref{fig:5vortices_observation_modes}(b) show, these higher modes are also more likely to be truncated in the rank reduction when the vortices are far from the sensors, due to these modes' relatively lower signal to noise ratio and higher corruption from spurious correlations. In contrast, the eigenvalues of the first three modes have comparable magnitude throughout the encounter.

\subsubsection{Five true vortices and three-vortex estimator}
\label{sec:5vortex-3estimate}
In the previous example, we assumed prior knowledge of the number of true vortices in the flow, but found that only a subset of directions of the vortex state vector could be updated at any instant. But there is no reason to expect that we know the number of true vortices {\em a priori}. It is enlightening to explore what happens when we estimate with fewer vortices than are present in the true flow, particularly because this is closer to a practical scenario, in which the true flow is a real (perhaps turbulent) viscous flow and we are using point vortices as a simplified representation of that flow. 

\begin{figure}
    \centering
    \begin{overpic}[width=1\linewidth]{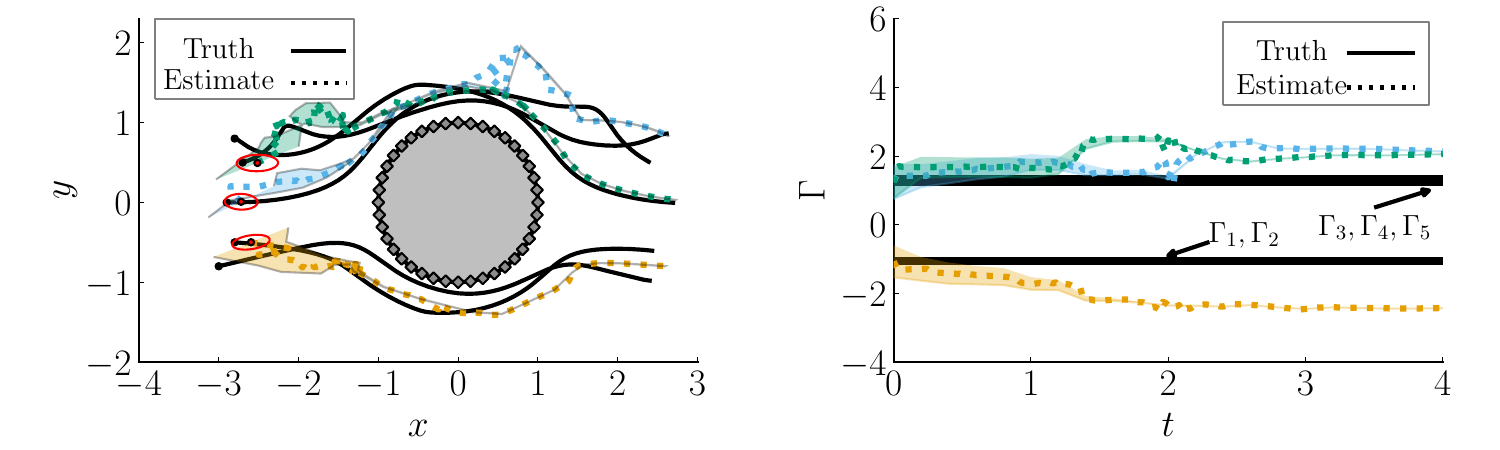}
    \put(-1,30){(a)}
    \put(53,30){(b)}
    \end{overpic}
    \caption{Five true vortices and three-vortex estimator. See Fig.~\ref{fig:5vortices_estimation} for legend.}
    \label{fig:3vortices_estimation}
\end{figure}

In all other aspects, the problem remains unchanged from the previous example, except the initial true states are slightly different, given by $(x^*_0/R_c,y^*_0/R_c,\Gamma/UR_c) = (-3,-0.8,-1)$, $(-2.8,-0.5,-1.1)$, $(-2.9,0,1.2)$, $(-2.7,0.5,1.3)$ and $(-2.8,0.8,1.4)$. Additionally, the dimension of the estimated state is now reduced to $\statedim = 9$, compared to 15 in the previous case. Figure~\ref{fig:3vortices_estimation}(a) shows the true and estimated trajectories. Obviously the three estimated vortices cannot precisely follow the trajectories of the five true ones. However, they do appear to estimate the cluster trajectories well. Indeed, Fig.~\ref{fig:3vortices_estimation}(b) confirms this clustering effect in the estimated vortex strengths: the estimated vortex nearest the two negatively signed true vortices below the cylinder is estimated to have a strength nearly equal to the sum of those two true vortices, denoted by $\Gamma_1$ and $\Gamma_2$. Similarly, the green and blue estimated vortices, which are positioned between the uppermost true vortices, $\Gamma_3$ through $\Gamma_5$, approximate the combined circulation of these three vortices.

\subsection{Comments on sequential estimation}

In \sect~\ref{sec:static_inference} we stressed the importance of a good prior for successful flow estimation. In sequential estimation we achieve this by leveraging the knowledge obtained in previous steps. To underscore this, we return to the example of da Silva and Colonius \cite{da2018ensemble}, in which they sought to estimate the grid-based velocity field of flows past various bodies with a modest number of sensors. Without a good prior, it is hopeless for the measurements from a few velocity sensors to correct the high-dimensional grid velocity. However, the forecast model (the Navier--Stokes equations and boundary conditions) is deterministic, so even if the initial ensemble is drawn from a poor distribution---e.g., a spatially-uncorrelated Gaussian distribution, or white noise---the state's ensemble is shaped over several forecast steps into a narrower and physically-plausible part of the state space. Indeed, in some of the examples in \cite{da2018ensemble}, the Navier-Stokes updates move the ensemble toward a trajectory on which the only uncertainty is in the phase of vortex shedding. Provided that there is sufficient initial variance of this phase among the ensemble members, the measurements are readily able to synchronize the evolving state onto the true phase. Even so, the benefits of this sequential improvement are still assisted by a good initial prior. Da Silva and Colonius \cite{da2018ensemble} found the best performance when the initial states in their ensemble of $\ensdim$ members were drawn from the space spanned by the leading $\ensdim$ proper orthogonal decomposition (POD) modes of the true flow. In the vortex estimation examples in this section, our initial prior consisted of non-overlapping Gaussian distributions for each estimator vortex, each of which was reasonably close to a true vortex. 

The last example in \sect~\ref{sec:vortex-sequence}, in which we interpreted measurements from a higher-dimensional system (five vortices) with a lower-dimensional representation (three vortices), provides an opportunity to make an important and more general point. The estimated trajectories in Fig.~\ref{fig:3vortices_estimation} show that the state covariance is quite small during most of the vortex--body interaction. However, it is crucial that we do not interpret this small covariance as precise knowledge of the true state. Rather, it represents a precise interpretation of the given measurements in the space in which we have confined our search. But that three-vortex space is a lower-dimensional manifold within the true space, and is thus inherently accompanied by a deterministic model error that we have no hope of estimating with our simplified state representation and observation model \cite{cohn1997introduction}. Fluid dynamics is replete with approximate representations of the flow field, each of which has a companion model error due to the restriction onto the simpler manifold, e.g., truncation error in grid-based CFD representations, subgrid-scale error in turbulence models, residual errors in reduced-order modal decompositions. For example, in the aforementioned study of da Silva and Colonius \cite{da2018ensemble}, in which they assembled their initial flow states from random combinations of POD modes, they might have instead chosen to work exclusively in the state space of POD coefficients. \newstuff{In that same spirit, in recent work \cite{mousavi2025sequentialestimationdisturbedaerodynamic}, the authors of this paper have performed stochastic estimation in a latent space learned from a neural-network autoencoder.} Some researchers have shown that it is possible to augment the state vector with the model error itself and, by virtue of a model that relates it to the measurements, estimate this error along with the simplified flow state \cite{da2020flow}.

Finally, we note that the LREnKF has provided a principled way to reduce the rank of the filtering so that the effect of spurious errors is mitigated, but there is a variety of other methods that improve the ensemble Kalman filter and we have only briefly listed a few. We also note that one could generalize the linear Kalman update \eqref{eq:kalman-analysis} to a nonlinear form, thereby relaxing the assumption of Gaussian distributions in the analysis step. Such is the goal of the nonlinear ensemble filter of Spantini et al.~\cite{spantini2022coupling}. Recall that the joint distribution $\probdist(\state,\meas)$ can be partitioned in two ways in equation~\eqref{eq:bayes0}: into the prior and likelihood, or into the marginal $\probdist(\meas)$ and the conditional $\probdist(\state|\meas)$ (i.e., the posterior, which we ultimately seek). Figure~\ref{fig:bayes-example} illustrated these partitions for a linear observation operator, and we made use of the nice properties of the Gaussian to extract the posterior. However, another way to extract this posterior is to map the joint distribution (the ellipse in Figure~\ref{fig:bayes-example}) into a standard normal distribution $\normaldist(0,\ident_{\statedim+\measdim})$ (a sphere of radius 1). The latter is trivial to partition into $\normaldist(0,\ident_{\measdim})\normaldist(0,\ident_{\statedim})$, and we can map the latter $\statedim$-dimensional part back to the state space for the desired posterior. In fact, this invertible mapping is always possible for {\em any} distribution, and this is the basic idea used in \cite{spantini2022coupling}. For this to be useful, we first have to have some knowledge of the joint distribution $\probdist(\state,\meas)$, and we can obtain such knowledge by assembling an ensemble of prior states with their associated predicted noisy measurements \eqref{eq:observenoise}. The method has been preliminarily evaluated in an aerodynamics problem by Le Provost et al.~\cite{le2021low}, but further investigation is needed to explore its merits in this context.




\section{Neural networks and other learned mappings}\label{sec:neural_network}
 
One of the points that we have tried to emphasize is that a good prior distribution is very helpful for flow estimation, since it can serve as a safety net when the rank of the measurement update is deficient or there is a multitude of possible candidates for the expected state. Underlying this pursuit of a good prior is the choice of the state space itself. If our desire is to estimate the flow's velocity field, then the obvious finite-dimensional space to work in is the space of all velocity fields on a grid or mesh. As we discussed in the previous section, our success in estimating flows directly in these high-dimensional spaces is predicated on a good initial prior and on the forecast model shepherding the ensemble into a narrower, physically-plausible region of the full space, where measurements can be more informative in the analysis updates.

Another approach we can take is to work explicitly in a low-dimensional state space, in which each point serves as a simplified representation of the full flow field. In many of the examples we have shown in this paper, we have represented the flow field in a state space describing the properties of a finite number of point vortices. Though this space is relatively low dimensional, physically interpretable, and has well known governing equations, point vortices also bring additional challenges due to the multitude of possible solutions that arise from their various symmetries, as we discussed in \sect~\ref{sec:vortex-static}. Instead, we could take a data-driven perspective: if we had a large set of data---sensor measurements and flow field snapshots---that roughly spans the expected behaviors of the flow, then we could potentially learn a state space and the associated operators from this data set, and use these to perform flow estimation. A familiar example of this approach is POD, in which the coefficients of the dominant POD modes of a flow form a reduced-dimensional space. To identify such low-dimensional spaces more generally, we turn to the extensive nonlinear function approximation toolbox of machine learning.

Deep neural networks (NN) and other tools from machine learning have provided a fertile ground in recent years for learning operators that estimate flow quantities from aerodynamic measurements. Zhong et al.~\cite{zhong2023sparse} have provided a pertinent example of this in the unsteady aerodynamics context. In that work, a small number of surface pressure measurements obtained from a CFD simulation of an airfoil in a disturbed flow were provided to a multi-layer perceptron (MLP) network. The network's resulting vector was a high-dimensional representation of the flow field (though still lower dimensional than the flow itself). This vector was then used by a convolutional neural network (CNN) to reconstruct the vorticity field or by another MLP to estimate the instantaneous lift. Morra et al.~\cite{morra2024ml} have recently shown that a trained deep operator network can be used in place of the full compressible Navier--Stokes equations to map upstream perturbations to downstream surface pressure measurements in a high-speed transitional boundary layer on a cone. They successfully used this approximate operator to estimate the amplitude of an upstream disturbance from downstream measurements taken from an experiment of the same geometry. Our objective in this paper is not to review the tools from machine learning and their application to fluid dynamics. Rather, our goal is to examine how to use these tools to support the stochastic flow estimation tasks that form the theme of this paper.

\begin{figure}
    \centering
    \includegraphics[width=1.0\textwidth, trim=0 0 0 0, clip]{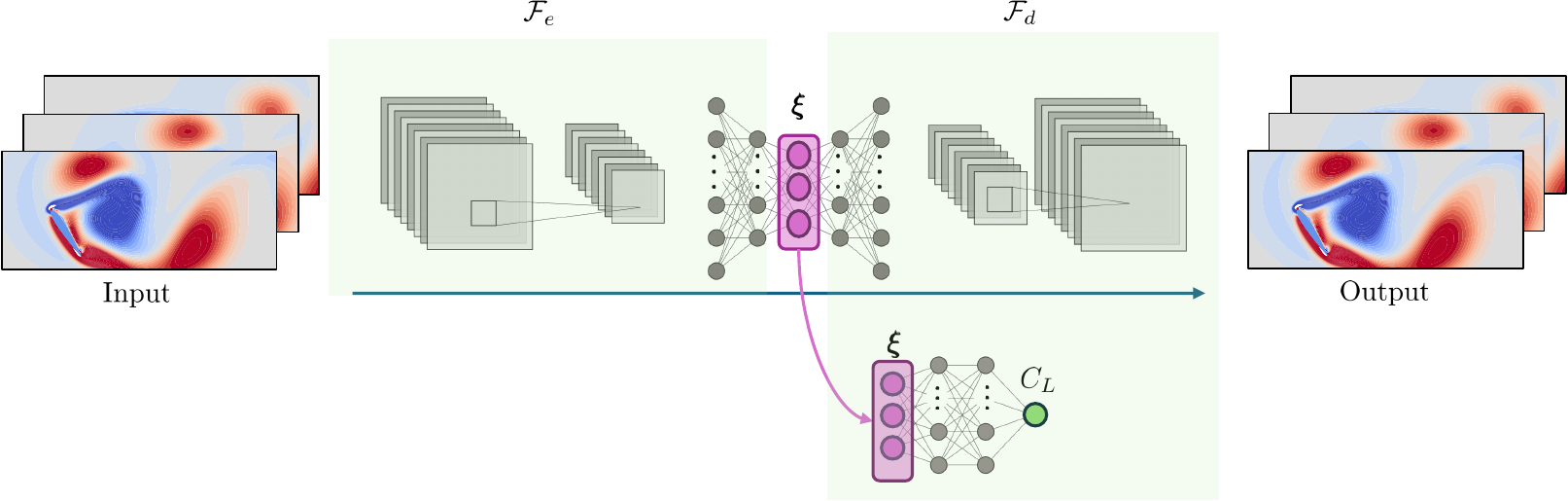}
    \caption{Lift-augmented autoencoder for non-linear reduced order of flow states.}
    \label{fig:autoencoder}
\end{figure}

Among current architectures of deep neural networks, the {\em autoencoder} provides the most powerful for distilling data into low-dimensional representations. During unsupervised training on a set of high-dimensional vectors, an autoencoder simultaneously learns to compress this data into a low-dimensional latent space and to expand latent vectors back to the high-dimensional space with minimal error. We specifically call attention to the lift-augmented autoencoder developed recently by Fukami and Taira \cite{fukami2023grasping} (and illustrated in Figure~\ref{fig:autoencoder}), in which the addition of a NN to predict lift helps to regularize an autoencoder that learns to compress vorticity field snapshots from a wide variety of undisturbed and disturbed two-dimensional flows past an airfoil at several angles of attack. The resulting latent space is three-dimensional, i.e., only three components are needed to uniquely specify the instantaneous flow field!

In other words, the autoencoder provides an invertible and differentiable mapping between the full flow state $\state$ and the reduced-dimensional latent state $\latent$, and for the purposes of our discussion, we will assume that such a mapping between $\state$ and $\latent$ exists. We can then perform our estimation exclusively in this latent space, training companion deep NN models to approximate the observation and forecasting tasks described in the previous sections: i.e., an observation operator, $\meas = \tilde{\observe}(\latent;\weights)$, with associated weights $w$, that maps latent vectors $\latent$ to predicted measurements $\meas$, and a forecast operator, $\latent_{\timeindex} = \tilde{\forecast}_\timeindex(\latent_{\timeindex-1};\weights)$, predicting the next latent state from the current one in a time sequence. (The unique mapping between $\state$ and $\latent$ preserves the Markovian character of the dynamics in the latent space.) Since our objective is to use these approximate operators with noisy measurements, we wish to consider their construction and use in our preferred probabilistic setting.

As with physics-based operators, our overarching objective with learned operators is to understand how uncertainty in the measurements $\meas$ translates into uncertainty in our state estimate $\state$---in this case, via uncertainty in the latent state $\latent$. If we combine the NN operators with models for the measurement and process noise, e.g., zero-mean Gaussians, then we have the ingredients of the sequential filtering equations---the transition probability \eqref{eq:transition} and the likelihood \eqref{eq:bayes-sequential}---and these enable us to track uncertainty through the flow prediction. \newstuff{This was the approach taken by Mousavi and Eldredge \cite{mousavi2025sequentialestimationdisturbedaerodynamic}, in which pressure measurements were used to estimate a seven-dimensional latent vector representation of the full flow state. The latent space and associated operators were learned from snapshots of the disturbed flow past a wing at several fixed angles of attack.} But there are important questions to explore before pursuing this: How does uncertainty propagate through the latent space? And how does uncertainty in the model itself enter the problem? In this section we will explore these questions by focusing on the approximate {\em inverse} observation operator, $\tilde{\invobserve}(\meas;\weights)$, which maps measurements $\meas$ to the latent state $\latent$. We will use this opportunity to quantify the various forms of uncertainty in this mapping.

\begin{figure}
    \centering
    \includegraphics[width=1.0\textwidth, trim=0 0 0 0, clip]{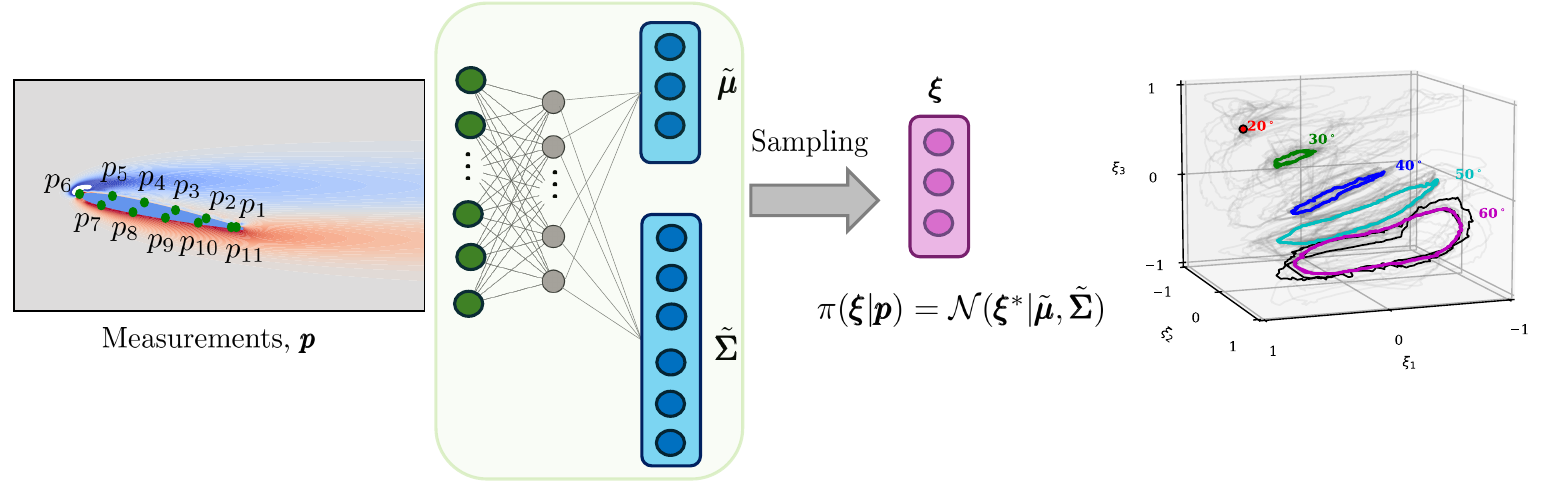}
    \caption{Neural network architecture for flow prediction from sparse noisy pressure measurements $\pmb{p}$, adapted with permission from Mousavi and Eldredge (J. Fluid Mech., 2025) \cite{mousavi2025loworderflowreconstructionuncertainty}. Baseline (undisturbed) trajectories (in color); disturbed trajectories (gray, with one case highlighted in black).}
    \label{fig:unc_prop_network}
\end{figure}

In our aerodynamics context, a representative example of this inverse mapping was investigated in \cite{mousavi2025loworderflowreconstructionuncertainty}, in which a lift-augmented autoencoder was trained on CFD simulation data from the same undisturbed and disturbed airfoil flows as in \cite{fukami2023grasping}. After training the autoencoder, the three-dimensional latent vector $\latent_\snapshot$ for each training snapshot $\snapshot$ was paired with a measurement vector, $\meas_\snapshot$, of the corresponding CFD-computed pressures at eleven evenly-spaced sensors on the surface of the airfoil. This latter vector also included the sensors' position coordinates to distinguish angles of attack and to respect the sensors' relative positioning. We will denote the training set of all such data pairs by $\data = \{\meas_\snapshot,\latent_\snapshot\}_{\snapshot=1}^{M}$.

The objectives we pursue here are to construct the mapping from measurement inputs $\meas$ to latent state outputs $\latent$ and to quantify the uncertainty of this mapping. There are a few ways that we could approach this. Perhaps the most straightforward is to use the data $\data$ to train a pointwise predictive (i.e., deterministic) NN, $\latent = \tilde{\invobserve}(\meas;\weights)$. We can then use a few different tools to characterize the uncertainty propagation through this network. The simplest is to use a Monte Carlo approach to characterize the network's uncertainty. We draw samples from an assumed measurement probability distribution, $\meas^i \sim \pi(\meas)$, e.g., by adding noise $\noisemember{i} \sim \normaldist(0,\noisecovar)$ to a snapshot $\meas_\snapshot$. The trained network propagates these samples to generate the corresponding latent space predictions, $\latent^i$. The resulting latent samples can then be modeled by a probability distribution, representing an approximation of the conditional probability $\pi(\latent|\meas)$.

However, because the neural network provides us with access to the Jacobian $\nabla_\meas \tilde{\invobserve}$ (through automatic differentiation), we can use this to characterize the uncertainty propagation more accurately. One such analytical method, \emph{stable distribution propagation}, was introduced by Petersen et al.~\cite{petersen2024uncertaintyquantificationstabledistribution} and is applicable to Gaussian and Cauchy-distributed uncertainties. The core idea is to approximate the network’s behavior via local linearization. In the context of $\tilde{\invobserve}$, the expected value and variance of the predicted output are estimated as 
\begin{equation}
\begin{aligned}
    \expect[\latent] &\approx \tilde{\invobserve}(\expect[\meas]), \\
    \text{Var}[\latent] &\approx \nabla_\meas \tilde{\invobserve} \cdot \text{Var}[\meas] \cdot (\nabla_\meas \tilde{\invobserve})^T,
\end{aligned}
\end{equation}
with $\nabla_\meas \tilde{\invobserve}$ denoting the Jacobian matrix of the operator $\tilde{\invobserve}$, which is evaluated at the expected value (mean) of a measurement. While this method provides a computationally efficient uncertainty quantification strategy, it is inherently limited in highly nonlinear networks, where local linearization fails to capture complex input-output mappings. To address this, Bibi et al.~\citep{bibi2018analytic} derived analytical, layer-wise expressions for propagating the first and second moments of input noise when the noise follows a Gaussian distribution.  It should also be noted that this local linearization is the same idea behind the state-space Gramian \eqref{eq:xgram} defined on the observation operator, and this can be easily extended to the inverse operator. The Gramian definition mitigates the effect of nonlinearity by computing expectations over the distribution of inputs, rather than only evaluating the Jacobian at the mean.

We could alternatively pursue an approach that combines the training and uncertainty quantification into a single task \cite{mousavi2025loworderflowreconstructionuncertainty}. To understand how this works, it is important to remember that training a mathematical model for least-squares regression can be viewed as finding a normal distribution $\normaldist(\tilde{\invobserve}(\meas;\weights),\covarbase)$ that maximizes the likelihood over the set of (random) training outputs \cite{bishop2006pattern}. We perform this maximization with respect to the weights $\weights$ of the network and the (as yet unknown) variance $\covarbase$ of the outputs. After training, the newly-determined $\covarbase$ is called the {\em aleatoric} (or data) uncertainty. When the inputs and outputs are both free of noise, as in the training of the deterministic network $\tilde{\invobserve}(\meas;\weights)$, then this perspective of regression is not particularly helpful. However, suppose we start with the same training data $\data$, but during each training epoch, to the inputs we add measurement noise drawn from $\normaldist(0,\noisecovar)$ ``on the fly''. We do not explicitly add noise to the outputs $\latent_\snapshot$, but by virtue of their pairing with random inputs we can also view them as random, as well. Training the network with this noisy data allows us to determine the aleatoric uncertainty of the outputs, and this uncertainty is attributable to the noisy inputs.

In fact, we can embrace this perspective further by expanding the network to predict both the mean {\em and} the variance as functions of the input, as shown in Fig.~\ref{fig:unc_prop_network}, adapted from \cite{mousavi2025loworderflowreconstructionuncertainty}. Such a neural network is called \emph{heteroscedastic} \cite{kendall2017uncertainties}, since the variance is not global, but dependent on the input. In our example, this input is pressure measurements, and the architecture illustrated in Fig.~\ref{fig:unc_prop_network} is designed to learn the parameters of a multivariate Gaussian distribution in the latent space,
\begin{equation}
\label{eq:heteroscedastic-nn}
    \normaldist(\tilde{\mu}(\meas;\weights),\tilde{\Sigma}(\meas;\weights)).
\end{equation}
Measurement noise is translated to the predicted latent vector by learning the parameters of this distribution. The right panel in Fig.~\ref{fig:unc_prop_network} depicts the low-dimensional latent space representation of disturbed flow fields.

The network is trained with the objective function defined as the negative log of the likelihood of the output training data, assuming this distribution.
The aleatoric uncertainty quantification in the predicted latent vector, arising from noise in surface pressure measurements, has been thoroughly investigated in \citep{mousavi2025loworderflowreconstructionuncertainty} using such a heteroscedastic network. An advantage of such a model is that it provides a generative model for the latent vectors for further Monte Carlo tasks. In particular, for a given measurement vector $\meas$, samples can be drawn from the normal distribution \eqref{eq:heteroscedastic-nn} and then propagated through the decoder and lift networks to quantify uncertainty in the flow field and the lift, respectively.

The methods discussed above represent common approaches for propagating data noise through NNs, thereby accounting for the data (or aleatoric) uncertainty in predictions. In these formulations, we assume that the mapping $\latent=\tilde{\invobserve}(\meas;\weights)$ is learned perfectly, meaning the NN parameters $\weights$ accurately capture the underlying true physical function. However, in practice, NNs often approximate complex dynamics with inherent modeling errors. These discrepancies arise due to limitations in the learned function, structural simplifications, or insufficient training data, leading to {\em epistemic} (or model) uncertainty--—the uncertainty associated with the model itself. Model uncertainty is particularly crucial in domains such as safety-critical control systems \citep{le2018uncertainty}, medical diagnosis \citep{laves2019quantifying}, and autonomous driving \citep{shafaei2018uncertainty}, where reliable uncertainty quantification is essential for risk-aware decision-making.

Taking the same inverse mapping $\latent=\tilde{g}(\meas;\weights)$ as discussed in the preceding paragraphs, the uncertainty in the model parameters $\weights$, denoted as $\pi(\weights|\data)$, can be estimated using various approaches. One widely used method is \emph{Deep Ensemble}, where a deterministic neural network is trained independently $N$ times, producing a set of models $\tilde{g}^i$, for $i=1,\cdots,N$. The predicted outputs from these independently trained models, $\{\latent^i\} = \{ \tilde{\invobserve}^i(\meas;\weights) \}_{i=1}^N$ can be interpreted as samples from the predictive distribution, providing an empirical estimation of epistemic uncertainty. This approach has been extensively used to enhance the robustness of neural networks \citep{ren2020deep, mcdermott2019deep}. However, its major limitation is computational inefficiency, as it requires training multiple networks independently—an impractical approach for large-scale or deep architectures where training costs are prohibitively high.

As in our previous discussions, a Bayesian framework provides a powerful means to quantify the probability distribution of NN parameters. By applying Bayes' theorem, the posterior distribution of the NN parameters is given by
\begin{equation}\label{eq:bayesNN}
    \pi (\weights|\data) = \frac{\pi(\data|\weights) \ \pi(\weights)}{\pi(\data)}
\end{equation}
where $\pi(\weights)$ represents the prior distribution over the parameters, and $\pi(\data|\weights)$, the likelihood, quantifies the probability of observing the training data given a specific parameter set $\weights$. This is just the aleatoric uncertainty discussed earlier. The denominator, $\pi(\data)$, normalizes the posterior and is obtained by integrating the joint distribution over all possible parameter values. This Bayesian formulation enables principled uncertainty quantification by capturing the full posterior distribution rather than relying on point estimates.

The intuition behind Eq.~\eqref{eq:bayesNN} lays the foundation for a transformative concept in the field of computer science: the Bayesian Neural Network (BNN). BNNs \citep{lampinen2001bayesian, titterington2004bayesian, goan2020bayesian} are stochastic NNs trained using Bayesian inference. In a BNN, the prior distribution over parameters can either be specified based on prior knowledge (such as from a pre-trained deterministic network \citep{krishnan2019efficient}) or learned during training. Given access to the posterior distribution, $\pi (\weights|\data)$, one can sample from it to approximate the predictive distribution, enabling principled uncertainty quantification and robust decision-making. This predictive distribution is computed as
\begin{equation}
    \pi(\latent|\meas,\data) = \int \pi(\latent|\meas,\weights) \ \pi(\weights|\data) \ \mathrm{d}\weights.
\end{equation}
However, obtaining the exact posterior distribution is challenging due to the intractability of the denominator in Eq.~\eqref{eq:bayesNN}. Instead of explicitly computing this integral, MCMC methods \citep{bardenet2017markov} provide a means of sampling from the true posterior distribution by leveraging the proportionality relationship: $\pi (\weights|\data) \propto \pi(\data|\weights) \ \pi(\weights)$. While MCMC is theoretically exact, it becomes computationally prohibitive for high-dimensional parameter spaces, particularly in deep NNs.

As a more scalable alternative, variational inference (VI) \citep{blei2017variational} approximates the true posterior $\pi (\weights|\data)$ with a tractable probability distribution $q_\phi(\weights)$, parameterized by $\phi$. The network is trained to optimize these parameters so that $q_\phi(\weights)$ closely resembles the true posterior. The quality of this approximation is measured using the Kullback-Leibler (KL) divergence \citep{kullback1951information}, which quantifies the discrepancy between the variational distribution and the true posterior:
\begin{equation}
    D_{KL} (q_\phi||\pi) = \int q_\phi(\weights) \ \log \left( \frac{q_\phi(\weights)}{\pi(\weights|\data)} \right)\,\mathrm{d}\weights.
\end{equation}
By minimizing the KL divergence, VI provides an efficient and scalable approach to approximate Bayesian inference in deep learning models.


Despite the effectiveness of VI in BNNs, directly learning the parameters of the approximate posterior remains computationally expensive and challenging to converge. To address this, Monte Carlo (MC) dropout, introduced by Gal and Ghahramani \citep{gal2016dropoutbayesianapproximationappendix, gal2016dropout}, provides an efficient approximation of the posterior distribution using dropout as a form of variational inference. They mathematically demonstrated that placing a random dropout layer after each dense layer is equivalent to performing VI in a deep Gaussian process. MC dropout is easy to implement and has been utilized by Mousavi and Eldredge \cite{mousavi2025loworderflowreconstructionuncertainty} to quantify the epistemic uncertainty in disturbed aerodynamics.



\section{Conclusions and future directions}\label{sec:conclusions}

In this paper, we have reviewed the tools of Bayesian inversion and sequential data assimilation, and framed this review in the context of estimating a fluid flow from the available sensors. In carrying out this task, we often have fewer sensors than states, or other degeneracies that lead to multiple solutions, rendering it ill-posed as an inversion problem. An underlying theme of this paper is that the Bayesian setting outfits us with crucial tools to deal with this ill-posedness: from a prior distribution of possible states, we seek a posterior distribution that is conditioned on the measurements.

The examples we have provided illustrate several key aspects of these tools. A spectral decomposition, either of the observation matrix or, more generally, the state- and measurement-space Gramians, provides a means of distinguishing the informative from the non-informative subspaces, so we can concentrate our state updates in the directions in which we have the highest trust in the available data. In contrast, the non-informative directions tell us about where our measurements have blind spots. The prior distribution is crucial for filling these blind spots. For example, we can use the prior to express known constraints on the flow state. In a sequential estimation, the prior is improved over time as we leverage past measurements and forecasts.

Sequential estimation effectively discovers a lower-order manifold on which the true flow is confined, on which measurements are more likely to be informative. The ensemble Kalman filter provides a straightforward and computationally efficient means of performing this sequential estimation. To ensure tractable computations, our inclination in fluid dynamics is to rely on small ensemble sizes. However, it is crucial in these small ensembles to prevent spurious correlations from corrupting the state updates. The spectral decomposition in the LREnKF provides one means of avoiding this, by discarding the untrustworthy directions. 

We can also improve the flow estimation task by using or learning a reduced-order representation of the state and working explicitly in this lower-dimensional latent space. Autoencoders provide a very effective neural network approach for identifying this latent space, and we can train other operators that map latent vectors to or from sensor measurements or forecast them forward in time. As with physics-based operators, it is crucial to understand how uncertainty is propagated through these learned operators. We have reviewed two types: aleatoric, or data, uncertainty, which quantifies how uncertainty in the data itself is quantified; and epistemic, or model, uncertainty, which expresses how the model's parameters influence the overall error.

We have only discussed a way of thinking about uncertainty in these learned operators, and there is still much more to do to exploit them in the sequential estimation framework. In this work, we have assumed that an autoencoder is trained from snapshots of the full flow field, free from noise, and we have treated the resulting latent vectors as a perfect surrogate of the full flow. Suppose, instead, we train an autoencoder from snapshots of planar PIV data in a three-dimensional flow past a wing. The latent space identified by this is inherently noisy and missing out-of-plane flow information. A variational autoencoder \cite{kingma2019introduction} would serve this task well. We might use the Bayesian framework to train an observation operator to assimilate both PIV data and measurements from pressure sensors distributed along both the span and chord of the wing, and exploit our knowledge of Navier--Stokes to fill in the remaining details of the three-dimensional flow around the wing.

There are a number of other interesting and important aspects of flow estimation that require further investigation. We have not addressed the important issue of sensor placement, though the probabilistic framework certainly provides a powerful toolbox for addressing this. The informative subspaces we have discussed here are obviously germane to sensors' observability over time. This review has not devoted any explicit consideration of the estimation of turbulent flows and their wide spectrum of length and time scales. It is important to note that some of the biggest advances to data assimilation of flows have been made in the area of numerical weather prediction, in which the breadth of these spectra is as large as any we encounter. Sparse sensors record the footprints of the largest scales of weather systems, and our knowledge of the governing equations of these flows---or, at least, our characterization of their statistics---enables us to rationalize these sensor measurements with wider knowledge of the flow structure. In linear stochastic estimation (LSE) \cite{adrian1975role,adrian1988stochastic}, developed by Adrian and co-workers, the mean velocity field in a turbulent flow is conditioned on events, i.e., pre-chosen characteristics of the velocity field at certain locations and times. The resulting ``conditional eddy'' structure is determined by the two-point correlations of the turbulent flow and provides a mean of inferring the expected behavior in the vicinity of certain observations. It should also added that the dynamics of turbulent flows have a much sharper sensitivity to the state than unsteady low Reynolds numbers. Data assimilation provides a means of mitigating this tendency toward spatio-temporal chaos. The recent review by Zaki \cite{zaki2024turbulence} provides an excellent discussion of this view and its mathematical underpinnings.

Flow and flight control are implicit motivations for much of what we have discussed in this paper, but we have not addressed the relationship between the estimated flow state and the actions that we take to achieve some control objective. In reinforcement learning, a machine learning framework that leverages experiential interactions with the environment to learn a control strategy, this strategy is typically expressed as a conditional probability, $\probdist(\mathsf{a}|\state)$---that is, given a system state, what is the distribution of actions $\mathsf{a}$ that we could take to maximize the expected rewards over time. One could reasonably ask, if we can successfully estimate $\state$ from observations $\meas$, then could we condition the actions on $\meas$ alone? To what extent do we explicitly need to estimate the state? These questions are critical for efficient training of successful reinforcement learning policies in fluid dynamics \cite{beckers2024deep}.



\section*{Acknowledgments}

The authors dedicate this paper to the loving memory of Dr.\ Mathieu Le Provost, who had already contributed so much to this topic in his short career. The authors also gratefully acknowledge the financial support provided by the National Science Foundation under award number 2247005.


\appendix

\section{Derivation of posterior mean and covariance in linear equations}\label{sec:appendix-linear}

Consider the linear observation model, $\meas = \observemat \state + \bias + \noise$, where $\bias$ is a constant vector, $\noise \sim \normaldist(0,\covar{\Noise})$, and $\state \sim \normaldist(\meanstate,\statecovar)$. Then the joint distribution of $\joint = (\state,\meas)$ is Gaussian, as we will show. The mean and covariance of this joint distribution can generically be written as
\begin{equation}
    \mean{\joint} = \begin{pmatrix}
        \meanstate \\ \meanmeas
    \end{pmatrix},\qquad \covar{\joint} = \begin{bmatrix}
        \covar{\state\state} & \covar{\state\meas} \\ \covar{\meas\state} & \covar{\meas\meas} 
    \end{bmatrix}.
\end{equation}
The entries in these can all be computed from expectations,
\begin{align}
    \meanstate &= \expect[\state] \label{eq:meanx}\\
    \meanmeas &= \expect[\meas] = \expect[\observemat\state + \bias + \noise] = \observemat\expect[\state] + \bias + \expect[\noise] = \observemat\meanstate + \bias \label{eq:meany}\\
    \covar{\state\state} &= \expect[(\state-\meanstate)(\state-\meanstate)^T] = \statecovar \label{eq:sigxx} \\
    \covar{\state\meas} &= \covar{\meas\state}^T = \expect[(\state-\meanstate)(\meas-\meanmeas)^T] = \expect[(\state-\meanstate) (\observemat(\state-\meanstate) + \noise)^T] \nonumber\\
    &= \expect[(\state-\meanstate) (\state-\meanstate)^T] \observemat^T + \expect[(\state-\meanstate)\noise^T] = \statecovar\observemat^T \\
    \covar{\meas\meas} &= \expect[(\meas-\meanmeas)(\meas-\meanmeas)^T] = \expect[(\observemat(\state-\meanstate) + \noise)(\observemat(\state-\meanstate) + \noise)^T] \nonumber\\
    &= \observemat\expect[(\state-\meanstate) (\state-\meanstate)^T] \observemat^T + \expect[\noise\noise^T] \nonumber\\ & \hspace{1cm} + \observemat\expect[(\state-\meanstate)\noise^T] + \expect[\noise(\state-\meanstate)^T] \observemat^T  \nonumber\\
    &= \observemat \statecovar \observemat^T + \covar{\Noise} \label{eq:sigyy}.
\end{align}
where we have used the fact that the state and the noise are uncorrelated.

If we take $\meas$ as the true state, $\truemeas$, then the joint distribution can be used to find the conditional distribution, which is also Gaussian. This is done by rewriting the log of the joint probability, a quadratic form for $\state$ and $\truemeas$,
\begin{equation}
    -\frac{1}{2} \begin{pmatrix}
        \state - \meanstate \\ \truemeas - \meanmeas
    \end{pmatrix}^T \begin{bmatrix}
        \covar{\state\state} & \covar{\state\meas} \\ \covar{\meas\state} & \covar{\meas\meas} 
    \end{bmatrix}^{-1} \begin{pmatrix}
        \state - \meanstate \\ \truemeas - \meanmeas
    \end{pmatrix}
\end{equation}
into a quadratic form for $\state$ only, treating $\truemeas$ as a constant \cite{bishop2006pattern}. The result is
\begin{equation}
    -\frac{1}{2} (\state - \mean{\state|\truemeas})^T \covar{\state|\truemeas}^{-1} (\state - \mean{\state|\truemeas}),
\end{equation}
where the conditional mean and covariance are
\begin{align}
    \mean{\state|\truemeas} &= \meanstate + \covar{\state\meas}\covar{\meas\meas}^{-1} (\truemeas - \meanmeas) \\
    \covar{\state|\truemeas} &= \statecovar - \covar{\state\meas}\covar{\meas\meas}^{-1} \covar{\meas\state}.
\end{align}
Substituting the expressions for $\meanmeas$, $\covar{\state\meas}$, and $\covar{\meas\meas}$, we arrive at the Kalman update equations,
\begin{align}
    \mean{\state|\truemeas} &= \meanstate + \kalman(\truemeas - \observemat\meanstate - \bias ) \\
    \covar{\state|\truemeas} &= (\ident - \kalman \observemat) \statecovar,
\end{align}
where
\begin{equation}
    \label{eq:kalman-app}
    \kalman = \covar{\state\meas}\covar{\meas\meas}^{-1} = \statecovar \observemat^T (\observemat \statecovar \observemat^T + \covar{\Noise})^{-1}.
\end{equation}

Using the whitened SVD of the observation matrix, $\observemat = \noisecovar^{1/2}USV^T\statecovar^{-1}$, it is easy to show that the covariances in \eqref{eq:sigxx}--\eqref{eq:sigyy} can be decomposed as follows: 
\begin{align}
    \label{eq:covar-svd}
    \statecovar &= \statecovar^{1/2} V \covar{\statesvd} V^T \statecovar^{1/2}, \qquad \covar{\statesvd} = \ident_\statedim, \\
    \covar{\state\meas} &= \statecovar^{1/2} V \covar{\statesvd\meassvd} U^T \noisecovar^{1/2}, \qquad \covar{\statesvd\meassvd} = S^T, \\
    \covar{\meas\meas} &= \noisecovar^{1/2}U \covar{\meassvd\meassvd} U^T\noisecovar^{1/2}, \qquad \covar{\meassvd\meassvd} = \ident_\measdim + SS^T.
\end{align}

\section{Ensemble approximations of mean and covariances}\label{sec:appendix-ensemble}

Suppose we have an ensemble $\{\statemember{\ensdex}\}_{\ensdex=1}^{\ensdim}$, composed of samples from some distribution, i.e. $\statemember{\ensdex} \sim \probdist_0(\state)$, and another ensemble of noise samples $\{\noisemember{\ensdex}\}_{\ensdex=1}^{\ensdim}$ from a zero-mean Gaussian distribution $\noisemember{\ensdex} \sim \normaldist(0,\noisecovar)$. Then the pairs $\{\statemember{\ensdex},\observe(\statemember{\ensdex})+\noisemember{\ensdex}\}_{\ensdex=1}^{\ensdim}$ represent an ensemble of samples from the joint distribution $\probdist(\state,\meas) = \normaldist(\meas|\observe(\state),\noisecovar) \probdist_0(\state)$. The following provide unbiased estimates of the means and covariances of this joint distribution: 
\begin{align}
\label{eq:ensxmean}
    \approxmean{\state} &= \frac{1}{\ensdim}\sum_{\ensdex=1}^{\ensdim} \statemember{\ensdex} \\
    \label{eq:ensymean}
    \approxmean{\meas} &=  \frac{1}{\ensdim}\sum_{\ensdex=1}^{\ensdim} \left(\observe(\statemember{\ensdex}) + \noisemember{\ensdex} \right) \\
    \label{eq:ensxxcov}
    \approxcovar{\state\state} &=  \frac{1}{\ensdim-1} \sum_{\ensdex=1}^{\ensdim} (\statemember{\ensdex}-\approxmean{\state})(\statemember{\ensdex}-\approxmean{\state})^T \\
    \label{eq:ensxycov}
    \approxcovar{\state\meas} &=  \frac{1}{\ensdim-1} \sum_{\ensdex=1}^{\ensdim} \left(\statemember{\ensdex}-\approxmean{\state}\right)\left(\observe(\statemember{\ensdex})+\noisemember{\ensdex}-\approxmean{\meas}\right)^T \\
\label{eq:ensyycov}
    \approxcovar{\meas\meas} &= \frac{1}{\ensdim-1} \sum_{\ensdex=1}^{\ensdim} \left(\observe(\statemember{\ensdex}) + \noisemember{\ensdex}-\approxmean{\meas}\right)\left(\observe(\statemember{\ensdex})+\noisemember{\ensdex}-\approxmean{\meas}\right)^T.
\end{align}
An important note about the ensemble covariance matrices \eqref{eq:ensxxcov}--\eqref{eq:ensyycov}: their rank is, at most, $\ensdim-1$. Why? Each term in the sums is a rank-1 matrix, since it projects a vector onto a space spanned by a single column vector, e.g., $\statemember{\ensdex}-\approxmean{\state}$. One might be tempted to think that the overall rank of the matrix is therefore $\ensdim$ (a sum of $\ensdim$ rank-1 matrices), but it is actually $\ensdim-1$, because these column vectors are not independent: they sum to zero, by virtue of the definitions of the means \eqref{eq:ensxmean} and \eqref{eq:ensymean}.

The Gramians \eqref{eq:xgram} and \eqref{eq:ygram} can be approximated by summing over the ensemble of samples $\{\statemember{\ensdex}\}_{\ensdex=1}^{\ensdim}$ drawn from the prior distribution,
\begin{equation}
\label{eq:xgram-approx}
    \hat{C}_\state = \frac{1}{\ensdim-1} \sum_{\ensdex=1}^{\ensdim}\left(\noisecovar^{-1/2}\nabla\observe(\statemember{\ensdex})\statecovar^{1/2}\right)^{T} \left(\noisecovar^{-1/2}\nabla\observe(\statemember{\ensdex})\statecovar^{1/2}\right)
\end{equation}
and 
\begin{equation}
\label{eq:ygram-approx}
    \hat{C}_\meas = \frac{1}{\ensdim-1} \sum_{\ensdex=1}^{\ensdim}\left(\noisecovar^{-1/2}\nabla\observe(\statemember{\ensdex})\statecovar^{1/2}\right) \left(\noisecovar^{-1/2}\nabla\observe(\statemember{\ensdex})\statecovar^{1/2}\right)^{T}.
\end{equation}

We can organize the members of the ensemble into the columns of a matrix, forming a $\Reals^{\statedim\times\ensdim}$ state ensemble
\begin{equation}
    \ensstate = \begin{bmatrix}
        \statemember{1} & \statemember{2} & \cdots & \statemember{M}
    \end{bmatrix},
\end{equation}
a $\Reals^{\measdim\times\ensdim}$ ensemble of predicted observations,
\begin{equation}
    \ensmeas = \begin{bmatrix}
        \observe(\statemember{1}) & \observe(\statemember{2})  & \cdots & \observe(\statemember{\ensdim})  
    \end{bmatrix},
\end{equation}
and $\Reals^{\measdim\times\ensdim}$ noise ensemble,
\begin{equation}
    \ensnoise = \begin{bmatrix}
        \noisemember{1} & \noisemember{2}  & \cdots & \noisemember{\ensdim} 
    \end{bmatrix}.
\end{equation}
Note that the noise-perturbed ensemble of observations is given by $\ensmeas + \ensnoise$. We also define the helpful vector $\ensones \in \Reals^{\ensdim}$, consisting of a column of ones, so that
\begin{equation}
    \approxmean{\state} = \frac{1}{\ensdim}\ensstate \ensones, \qquad \approxmean{\meas} = \frac{1}{\ensdim}\ensmeas \ensones.
\end{equation}

From these, we can define the anomaly matrices, which remove the mean from the ensemble and divide by $\sqrt{\ensdim-1}$, e.g.,
\begin{equation}
    \ensstate' = \frac{1}{\sqrt{\ensdim-1}}\left( \ensstate - \approxmean{\state} \ensones^T\right) = \frac{1}{\sqrt{\ensdim-1}}\ensstate\left(\ident_{\ensdim}  - \frac{1}{\ensdim}\ensones\ensones^T\right),
\end{equation}
where $\ident_\ensdim$ denotes the identity matrix in $\ensdim$ dimensions. From the anomaly matrices, the approximate covariances are easily obtained, using the noise-perturbed observations, 
\begin{equation}
\label{eq:enscovar-matrix}
    \approxcovar{\state\state} = \ensstate' (\ensstate')^T, \quad  \approxcovar{\state\meas} = \ensstate' (\ensmeas')^T, \quad \approxcovar{\meas\meas} = \ensmeas' (\ensmeas')^T + \ensnoise' (\ensnoise')^T.
\end{equation}

It is important to note that, in writing these approximations of the true covariances for an infinite ensemble, we have omitted cross-covariances between the states or predicted observations with the noise, $\ensstate' (\ensnoise')^T$ and $\ensmeas' (\ensnoise')^T$. These cross-covariances are zero in the limit of infinite ensemble size, but for a finite-sized ensemble they are not. There are also likely to be spurious correlations among the components of $\state$ in \eqref{eq:ensxxcov}.

Given a set of eigenvectors $V_{\staterank}$ and $U_{\measrank}$ of $C_\state$ and $C_\meas$ and reduced to ranks $\staterank\leq \statedim$ and $\measrank\leq \measdim$, respectively, we can compute new coordinates for the state and measurement spaces as
\begin{equation}
    \statesvd = V_{\staterank}^T\statecovar^{-1/2} (\state - \approxmean{\state}), \qquad \meassvd = U_{\measrank}^T\noisecovar^{-1/2} (\meas - \approxmean{\meas}).
\end{equation}
This can be done for any member of an ensemble, leading naturally to transformed anomaly matrix definitions
\begin{align}
    \ensstatesvd &= V_{\staterank}^T\statecovar^{-1/2} \ensstate', \\
    \ensmeassvd &= U_{\measrank}^T\noisecovar^{-1/2} \ensmeas', \\
    \ensnoisesvd &= U_{\measrank}^T\noisecovar^{-1/2} \ensnoise'.
\end{align}
From these, we can compute approximations of the covariances in these transformed coordinates,
\begin{equation}
\label{eq:enscovar-reduced}
    \approxcovar{\statesvd\statesvd} =  \ensstatesvd \ensstatesvd^T, \quad \approxcovar{\statesvd\meassvd} =  \ensstatesvd \ensmeassvd^T, \quad \approxcovar{\meassvd\meassvd} =  \ensmeassvd \ensmeassvd^T + \ensnoisesvd \ensnoisesvd^T.
\end{equation}

\bibliography{refs}

\end{document}